\documentclass[showpacs,nofootinbib,preprintnumbers,prd,superscriptaddress, twocolumn]{revtex4}

\usepackage{amsmath}
\usepackage{amsfonts}
\usepackage{amssymb}
\usepackage{graphicx}
\usepackage[titletoc]{appendix}
\usepackage{color}
\usepackage{hyperref}
\usepackage{cleveref}
\usepackage[rightcaption]{sidecap}
\usepackage{subfigure}
\usepackage{comment}

\usepackage{dcolumn}

\usepackage{array}
\usepackage{ctable}
\usepackage{multirow}
\usepackage{siunitx}
\usepackage{longtable}
\usepackage{tabularx}
\usepackage{booktabs,ragged2e}

\newcolumntype{R}[1]{>{\RaggedLeft\arraybackslash}p{#1}}

\graphicspath{{Graphics/}}

\def\be{\begin{equation}}
\def\ee{\end{equation}}
\def\bea{\begin{eqnarray}}
\def\eea{\end{eqnarray}}

\definecolor{vividviolet}{rgb}{0.62, 0.0, 1.0}
\definecolor{amaranth}{rgb}{0.9, 0.17, 0.31}
\definecolor{palatinateblue}{rgb}{0.15, 0.23, 0.89}
\definecolor{brightpink}{rgb}{1.0, 0.0, 0.5}
\definecolor{cornflowerblue}{rgb}{0.39, 0.58, 0.93}
\definecolor{deepcarminepink}{rgb}{0.94, 0.19, 0.22}
\definecolor{radicalred}{rgb}{1.0, 0.21, 0.37}
\hypersetup{ linktoc=all,
    colorlinks, linkcolor={palatinateblue},
    citecolor={brightpink}, urlcolor={amaranth}
}

\begin{document}

\title{Alleviating the cosmological constant problem from  particle production}

\author{Alessio Belfiglio}
\email{alessio.belfiglio@unicam.it}
\affiliation{Universit\`a di Camerino, Divisione di Fisica, Via Madonna delle carceri 9, 62032 Camerino, Italy.}
\affiliation{Istituto Nazionale di Fisica Nucleare, Sez. di Perugia, Via A. Pascoli 23c, 06123 Perugia, Italy.}

\author{Roberto Giamb\`o}
\email{roberto.giambo@unicam.it}
\affiliation{Universit\`a di Camerino, Divisione di Fisica, Via Madonna delle carceri 9, 62032 Camerino, Italy.}
\affiliation{Istituto Nazionale di Fisica Nucleare, Sez. di Perugia, Via A. Pascoli 23c, 06123 Perugia, Italy.}

\author{Orlando Luongo}
\email{orlando.luongo@unicam.it}
\affiliation{Universit\`a di Camerino, Divisione di Fisica, Via Madonna delle carceri 9, 62032 Camerino, Italy.}
\affiliation{Dipartimento di Matematica, Universit\`a di Pisa, Largo B. Pontecorvo 5, 56127 Pisa, Italy.}
\affiliation{NNLOT, Al-Farabi Kazakh National University, Al-Farabi av. 71, 050040 Almaty, Kazakhstan.}

\begin{abstract}
We explore a toy model mechanism of geometric cancellation, alleviating the (classical) cosmological constant problem. To do so, we assume at primordial times that vacuum energy fuels an inflationary  quadratic hilltop potential nonminimally coupled to gravity through a standard Yukawa-like interacting term, whose background lies on a perturbed Friedmann-Robertson-Walker metric. We demonstrate how vacuum energy release transforms into geometric particles,  adopting a quasi-de Sitter phase where we compute the expected particle density and mass ranges.  Perturbations are introduced by means of the usual external-field appproximation, so that the back-reaction of the created particles on the geometry is not considered here. We discuss the limitations of this approach and we also suggest possible refinements. We then propose the most suitable dark matter candidates, showing under which circumstances we can interpret dark matter as constituted by geometric quasiparticles. We confront our predictions with quantum particle production and constraints made using a Higgs portal. In addition, the role of the bare cosmological constant is reinterpreted to speed up the universe today. Thus, consequences on the standard $\Lambda$CDM paradigm are critically highlighted, showing how both coincidence and fine-tuning issues can be healed requiring the Israel-Darmois matching conditions between our involved inhomogeneous and homogeneous phases. 
\end{abstract}



\maketitle

\section{Overview}\label{introduzione}

The cosmological constant problem is the undeniable tension in reconciling the observed values of vacuum energy density  and theoretical large value of zero-point quantum vacuum fluctuations\footnote{According to the standard lore of quantum field theory, ground state energy supports non-zero excitations as both potential and  kinetic energies cannot vanish at the same time, providing  extremely large quantum fluctuations.} \cite{wein1}. This issue affects theoretical  physics and its resolution would certainly convey a very important step towards understanding physics beyond current standard models of cosmology and particle physics \cite{martin}. The corresponding background cosmology, namely the $\Lambda$CDM model \cite{pad,ratra}, associated to the standard Big Bang scenario, is jeopardized by fine-tuning and coincidence issues as consequence of the aforementioned cosmological constant problem \cite{copeland}. Thus, it is likely that solving the latter would justify the exact dark energy magnitude, exhibiting a self-consistent scheme for late-time cosmology. 

On the other side, early-time cosmology is driven by a widely-established inflationary epoch where the universe speeds up under the action of an inflaton field \cite{inflation}. Commonly, it is believed the current accelerated phase and inflation represent different scenarios, despite models unifying both the two epochs  are currently subject of intensive studies, see e.g. \cite{DLM,DiElle} and references therein.

In this work, we propose a toy model that tries to partially heal the  (classical) cosmological constant problem\footnote{The cosmological constant problem is often split in \emph{classical} and \emph{quantum}. Here, we focus on the first case only. For additional details, one can see Ref. \cite{martin}. In this work, we implicitly refer to the cosmological constant problem as its classical version only.}. In particular, we couple the inflaton field with curvature, we can obtain the inflationary dynamics and a  particle production induced by curvature that we may interpret as dark matter. We conjecture that the corresponding magnitude of such particles may cancel out the degrees of freedom of vacuum energy, counterbalancing its value and \emph{de facto} alleviating the huge discrepancy between observations and predictions. To do so, we propose a suitable value for the bare cosmological constant today, assuming it to drive the universe at current time. To do so, following the Sakharov hypothesis \cite{sakharov}, stating that the stress-energy tensor  of a field placed in the vacuum state must be proportional to the constant\footnote{This can be shown starting from flat (Minkowski) spacetime, where the only invariant tensor is $\eta_{\mu \nu}$. Since the vacuum state must be the same for all observers, this implies $\langle 0 \lvert T_{\mu \nu} \rvert 0 \rangle \propto \eta_{\mu \nu}$. Moving to curved spacetime, conservation of stress-energy tensor requires $\langle T_{\mu \nu} \rangle=-\rho_{\rm vac} g_{\mu \nu}$, with $\rho_{\rm vac}$ constant. In the following we will then write the cosmological constant as $\Lambda= \Lambda_B + \rho_{\rm vac}$, where $\Lambda_B$ is the bare cosmological constant driving current expansion of the universe. For a different perspective see, for example, \cite{varyL}.} vacuum energy density $\rho_{\rm vac}$, we argue that the so-produced dark matter particles are forced to be weakly-interacting and stable. We discuss their properties and  assume that they could be under the form of quasiparticles in agreement with previous findings, see e.g. \cite{BLM}. In so doing, we show that a passage from an initial quasi-de Sitter phase in a perturbed Friedmann-Robertson-Walker (FRW) spacetime to a radiation dominated universe is needful. If so, passing through these two phases, \textit{i.e.}, from a inhomogeneous to homogeneous universe, would imply two main processes: 1) inflation first, driven by an effective curvature-coupled inflaton potential and 2)  dark matter production fueled by  vacuum energy release and due to the coupling with geometry. In our treatment, we neglect possible \emph{back-reaction} mechanisms, i.e., we do not show how particle production acts back on the spacetime geometry, thus modifying the original perturbations\footnote{As discussed in \cite{referenza2}, some arguments suggest that the reaction of particle creation back on the gravitational field would modify the expansion, reducing the creation rate. A lower creation rate does not affect in any critical way our model, it simply implies a longer time to produce the desired number density of particles, as we will see. However, it is evident that a fully self-consistent treatment of gravitational particle production needs to properly address the issue of back-reaction. We will come back to this point later in the text.}. We also discuss under which circumstances quantum mechanisms of particle production could be sudominant than geometric particle production. Further, we show suitable intervals of mass ranges for our dark matter candidates and we compare our expectations with suitable examples of Higgs portal. Moreover, we discuss heuristically both the fine-tuning and coincidence problems by adopting the Israel-Darmois conditions to connect our inhomogeneous and homogeneous universes. In this respect, we conjecture the origin of the bare cosmological constant as due to matter pressure only, in agreement with a mechanism of vacuum energy cancellation recently proposed in Refs. \cite{LM,DLM}. Finally, consequences  on the $\Lambda$CDM paradigm are investigated.

The paper is outlined as follows. In Section \ref{sez2} we propose an effective potential driving inflation, carrying vacuum energy that couples to gravity and we discuss its implications in both inflation and particle production. The latter is well-described by using a perturbed FRW  to get particle contributions from vacuum energy. Consequences after inflation, namely in the reheating, radiation and matter eras, are investigated. The predictions of our dark matter constituents are reported in Sec. \ref{sez2E}. The consequences at late-time, about the coincidence and fine-tuning caveats are highlighted in Sec. \ref{sez2F}. The role of the bare cosmological constant is also debated. Theoretical consequences of our recipe have been moreover discussed in Section \ref{sez3}, emphasizing the strengths and limitations of our model. The role of quantum particles has been reviewed. Excluded ranges of masses for our geometric dark matter particles have been also discussed. Conclusions and perspectives are drawn in Section \ref{sez4}. Appendices concerning the details of our computations have been also shown at the end of our manuscript.


\section{Lagrangian setup} \label{sez2}

In this section, we  investigate particle production that occurs as the universe undergoes a perturbed phase, \textit{i.e.}, where it turns out to be not-perfectly homogeneous and isotropic. To justify this fact in view of the cosmological principle, we will assume as basic demand, widely-considered in the literature \cite{inflation, referenza4}, that metric perturbations originate from quantum fluctuations of the inflaton field throughout all the inflationary phase. Thus, inflation generates quantum fluctuations responsible for producing particles at primordial times \cite{inflaparticle}. 

We work out the latter ansatz to investigate whether particles inferred from geometry only can influence the overall dynamics at primordial times. In fact, we are excluding possible couplings of our fields with other fields from the standard model of particle physics. Moreover, we are also neglecting “quantum" particle production from vacuum\footnote{With this expression we refer to gravitational particle production due to the sole expansion of spacetime, see e.g. the seminal work \cite{par} or \cite{martinm, for} for more recent reviews. Such particle production has a quantum nature, since it relies on the fact that the initial (quantum) vacuum state is no longer seen as a vacuum as the universe expands, thus leading to particle creation. More details are given in Sec. \ref{sez3}.}, that would imply particle pairs that in principle could annihilate. We will come back to this issue later on.

At primordial times, therefore, the universe is clearly dominated by vacuum energy \cite{vac1} that, by construction, tends to highly accelerate the universe \cite{vac2}. The effect of particle production would reduce the net amount of vacuum energy, breaking  the universe down. To model this process, we choose a potential that carries out vacuum energy with it throughout inflation, whose scalar field is naively associated to inflaton\footnote{In principle, other approaches are possible. In the case of Higgs field, for instance, one gets the Higgs inflation \cite{higgsinflation}. Alternative views are also related to \cite{alter}.}. 

To simplify our scheme, we compute geometric particle production as due to inhomogeneities over a perturbed FRW background\footnote{As a matter of fact, one can imagine to change the spacetime instead of perturbing FRW background. However, this would imply to know \emph{a priori} the underlying metric that is clearly unknown.}. To account the high acceleration due to inflationary epoch, we assume that particles are produced during an approximate de Sitter phase, \textit{i.e.}, having a fast-evolving scale factor. Undoubtedly, in a pure de Sitter phase we cannot escape from accelerating the universe. Consequently, postulating a suitable version of our scalar-field potential is crucial in order to get a graceful exit from inflation, as we will see.

A scalar field Lagrangian is therefore introduced, as composed by three main parts, namely $\mathcal L=\mathcal L_1+\mathcal L_2+\mathcal L_3$ that read
\begin{align}\label{eq1}
\mathcal{L}_1&= \frac{1}{2}  g^{\mu \nu}\phi_{, \mu} \phi_{,\nu}\,\quad[{\rm Kinetic\,\, term}]\nonumber\\
\mathcal{L}_2&=-\frac{\xi}{2} R \phi^2\,\,\,\,\qquad[{\rm Yukawa-like \,\,term}]\nonumber\\
\mathcal{L}_3&=-V(\phi)\,\,\,\,\,\,\,\,\,\,\,\,\quad[{\rm Potential\,\,term}]
\end{align} 
whose physical meanings are reported in the square brackets on the right, with the minus sign for $\mathcal L_2$ and $\mathcal L_3$ imposed adopting the signature convention $(+, -, -, -)$ for $\mathcal L$. 

Choosing the Yukawa interaction implies to couple the gravity sector to the scalar field $\phi$. The interaction that we chose turns out to be the simplest non-minimal contribution to the Lagrangian. Simpler approaches, namely minimal couplings, would not produce  remarkable results in view of particle production. 

Finally, the coupling constant $\xi$ implies non-minimal coupling with curvature that resembles a Yukawa-like interaction between the scalar field $\phi$ and curvature itself, \textit{i.e.}, showing an illuminating toy model describing self-interacting fields with spacetime\footnote{Usually  Yukawa interaction involves three complex fields with a  coupling constant  associated to one of the four fundamental forces. Since this is not exactly our case, we here use \emph{Yukawa-like}. }.

In our scheme, $V(\phi)$ is the inflationary potential that drives the universe to accelerate during inflation. Consequently, the field $\phi$ corresponds to the inflaton, that in our model is thought to evolve from small to large field excitations, with small curvature at the end of inflation. 

The Yukawa-like term carries with it the interaction, so as in particle physics one can imagine \emph{to dress the field} $\phi$ with the interaction itself \cite{quasi}. Consequently, since the interaction involves curvature the corresponding particles would be quasiparticles, interpreted as   excitations  between geometry and inflaton. This hypothesis discussed in Refs. \cite{BLM,niki} resembles free standard particles, but provides for them a different mass and, more in general, different physical properties. As we clarify later, we interpret those particles, produced within our landscape, as dark matter candidates. The mechanism of geometric particle production is clearly due to the kind of coupling between the inflaton and the Ricci scalar and agrees with previous approaches that seem to provide similar outcomes \cite{niki}. Rephrasing this concept, we here propose geometric particles of dark matter within the context of pure general relativity\footnote{Actually, the concept of geometric particles, or better geometric modes, of dark matter is also associated to extended theories of gravity. There, geometric particles are associated to higher orders of corrections within the Hilbert-Einstein's action \cite{CDL}.} (GR). 

Last but not least, we conventionally describe the  universe evolution in terms of conformal time\footnote{See Appendix \ref{appA} for a brief introduction to the notion of conformal time in cosmology and its relation to cosmic time.}, $\tau$, having the conformally-flat FRW metric to be 

\begin{equation}
g_{\mu \nu}=a^2(\tau) \eta_{\mu \nu}\,,
\end{equation}
with $\eta_{\mu\nu}$ the standard Minkowski metric.  The ansatz for the scale factor in the various epochs considered, and the corresponding matching conditions, will be discussed later in the text.

\subsection{Inflationary potential} \label{sez2A}

Adopting Eq. \eqref{eq1}, we do not know \emph{a priori} the most suitable choice for the potential. Following Planck satellite results \cite{referenza3}, there is no consensus about the most suitable scalar field potential that drives inflation. The corresponding experimental results provided several approaches that are still valid, ruling out other versions of $V(\phi)$. Among all the most promising possibilities, the hilltop potentials have not been excluded yet \cite{kad,boub} and may well-adapt to our scopes of healing the  longstanding cosmological constant problem, producing \emph{de facto} particles from quantum vacuum energy. Indeed, the choice
\begin{equation}
    V(\phi)= \Lambda^4 (1-\phi^n/\mu_n^n+ \dots),
\end{equation}
with $n=2;4$, involves a typically-large early cosmological constant, which may drive cosmological inflation. Even though assuming hilltop potentials is not the unique possibility, it appears as a remarkable toy approach that considers the presence of a potential driving inflation with vacuum energy and permits one to analytically integrate the subsequent equations related to particle production amount. Further, such potential has the advantage of driving inflation for small fields, $\phi\simeq0$, thus leaving the constant cosmological term to be responsible for a large scalar curvature and particle production. More complicated models can also be invoked, e.g. by assuming large field approaches, like the Starobinski potential,  albeit in this case the amount of particles would be mostly due to the interaction between inflaton and curvature, thus complicating the overall treatment.

Hence, to guarantee the above prescriptions to hold, our strategy consists in the following two steps:

\begin{itemize}
    \item[--] we split the universe into different epochs. The first is dominated by the inflaton field. The subsequent describes reheating and afterwards radiation dominated epoch until our era, \textit{i.e.}, late-time. During inflation, we write FRW perturbations within the de Sitter spacetime as generated by quantum fluctuations of the inflaton field;  
    \item[--] we evaluate geometric particle production \cite{referenza2, cesp} during inflation, adopting the simplest choice for the coupling constant $\xi$, namely the conformal coupling\footnote{In Appendix \ref{appAA} we generalize our approach to the case of an unspecified coupling, deriving the corresponding solutions for the field $\phi$.} $\xi=1/6$. To do so, we focus on geometry to fuel particle production, neglecting the quantum particle production related to the Bogoliubov coefficients \cite{park,referenza9}, as above remarked.
\end{itemize}
To work out our treatments, from Eq. \eqref{eq1}, the free equation of motion for $\phi$ reads
\be \label{eq3}
\square_\eta \chi+ V^\prime(\phi \rightarrow \chi/a)\ a^3=0,
\ee
with $\square_\eta\equiv \partial_\mu\partial^\mu$ in conformal time and  $V'(\phi) \equiv \partial V/\partial \phi$. From Eq. \eqref{eq3}, since we rescaled the field itself by the ansatz
\be \label{eq4}
\phi(x)= \chi(x)/a,
\ee
the friction term, namely $\sim \dot \phi H$, disappears as a natural consequence of our choice, as well-known in the literature, see e.g. \cite{dun}.

Since our geometric particle production occurs at early stages of inflationary domain, namely as $\phi$ is small, the case $n=4$ is disfavored to describe our prescription than $n=2$, that also has the advantage to provide analytical dynamical solutions in strict analogy to the case of chaotic potential $V(\phi)=m^2 \phi^2/2$. Accordingly, we write the \emph{hilltop quadratic} potential by 
\be \label{eq5}
V(\phi)= \Lambda^4 (1-\phi^2/\mu_2^2+ \dots),
\ee
where $\Lambda^4$ corresponds to the vacuum energy density during inflation \cite{dodels, rio2}. 

The above potential is defined independently from the shift $V(\phi)\rightarrow V(\phi)+\mathcal C$, with $\mathcal C$ a generic constant, by simply rescaling the values of $\Lambda^4$ and $\mu_2^2$. This guarantees that, modifying the potential by adding a constant, the cosmological constant problem is not restored.  The scale $\mu_2$ is intimately related to the field width, \textit{i.e.}, to the field variation during inflation. 

\subsection{Effective coupling with geometry}

Inflation occurs as $\phi\simeq0$ and, by virtue of Eq. \eqref{eq1}, we define the corresponding effective potential driving our inflationary phase as
\begin{equation}\label{effectivepotential}
    V^{{\rm eff}}(\phi,R)\equiv \Lambda^4 (1-\phi^2/\mu_2^2)+\xi R\phi^2\,,
\end{equation}
having constructed the sum of both hilltop potential and geometric coupling without any more complicated interactions. During inflation we can approximate it for small fields, having \emph{de facto} that it can reduce to a slightly evolving vacuum energy contribution $\sim \Lambda^4$. 

Clearly the dynamics of Eq. \eqref{effectivepotential} is not fully-stable as due to the typology of coupling with scalar curvature, here the Yukawa-like one. In particular, once the original hilltop potential is modified through the geometric coupling, it is possible \emph{a priori} not to get a \emph{graceful exit}. In principle, we are here proposing a toy model where the coupling with curvature can play the role of producing particles, but further investigations on Eq. \eqref{effectivepotential}, to work out how inflation naturally ends, are essential. In other words, one has to investigate which kind of more complicated curvature coupling may be included into the above scenario, in order to exit inflation. Limiting to this toy model, we make some heuristic considerations on how inflation may end later in the manuscript. 

At this stage, plugging  Eq. \eqref{eq5} into \eqref{eq3}, we get  
\be \label{eq6}
\square_\eta \chi - 2 a^2 \frac{\Lambda^4}{\mu_2^2} \chi=0.
\ee
Here, Eq. \eqref{eq6} can be analytically solved to adapt throughout inflation occurs. We focus on two phases below, namely during and after inflation. We thus analyze how to produce particles and how to interpret them as dark matter, and then we discuss the consequences of our recipe immediately after inflation, up to our times. 


\subsection{Phase A: Starting with the inflationary stage} \label{sez2B}

Theoretically speaking, inflation lasts inside $-\infty < \tau < 0$. Around $\tau\simeq0$, \textit{i.e.}, as inflation ends, a de Sitter phase would naturally diverge and consequently unphysical divergences could occur. To avoid such singularities, the scale factor can be rewritten as prompted in Ref.  \cite{referenza7}:
\be \label{eq43}
a(\tau)= \frac{1}{1-H_I\tau},\ \ \ \ \tau \leq 0,
\ee
where we conventionally baptize the Hubble constant  with  $H_I$, during inflation. It is evident that Eq. \eqref{eq43} does not provide any pathology within the range $-\infty < \tau \leq 0$, letting our model to work better during and immediately after inflation.

For the sake of clearness, it is well-established that during inflationary stage the Hubble rate is not \emph{exactly} a constant. It may slightly change with time, leading to a \emph{quasi-de Sitter expansion}. Assuming a de Sitter phase is therefore an approximation that, however, works well in describing the overall evolution of the scalar field during inflation. We will come back to analyze this issue later throughout the text.

\subsubsection{Dynamical solutions}

Now, bearing the ansatz \eqref{eq43} in mind, Eq. \eqref{eq6} gives 
\be \label{eq44}
\square_\eta \chi- \frac{2}{(1-H_I\tau)^2} \frac{\Lambda^4}{\mu_2^2}\chi=0,
\ee
whose general solution can be recast by
\be \label{eq9}
\chi({\bf x}, \tau)= \frac{f_k(\tau) e^{i {\bf k} \cdot {\bf x}}}{(2 \pi)^{3/2}}.
\ee
Here, the field modes $f_k$ satisfy the differential equation 
\be \label{eq45}
\ddot{f}_k+ \bigg[k^2- \frac{1}{(1-H_I\tau)^2} \left(\frac{2\Lambda^4}{\mu_2^2} \right)  \bigg]f_k=0.
\ee
Hence, Eq. \eqref{eq45} can be more compactly written as
\be \label{eq46}
\ddot{f}_k+ \bigg[k^2- \frac{1}{\tilde{\tau}^2} \left(\frac{2\Lambda^4}{\mu_2^2H_I^2} \right)  \bigg]f_k=0,
\ee
having introduced the new variable $\tilde{\tau}=\tau-1/H_I$.  This equation has the form
\be \label{Hankeq}
\ddot{f}_k+ \left[ k^2- \frac{1}{\tilde{\tau}^2} \left(\nu^2-\frac{1}{4} \right)   \right] f_k=0,
\ee
admitting general solutions given in terms of Hankel's functions \cite{referenza4}
\be \label{Hankesol}
f_k(\tau)= \sqrt{-\tilde{\tau}}\left[ c_1(k) H^{(1)}_\nu(-k\tilde{\tau})+ c_2(k) H^{(2)}_\nu(-k\tilde{\tau})   \right].
\ee

\subsubsection{Bunch-Davies state for vacuum}

The constants $c_1(k)$ and $c_2(k)$ are determined by selecting the vacuum state in the de Sitter space. As it is well-known from quantum field theory, a general curved spacetime does not admit a canonical, or even preferred, vacuum state \cite{referenza1}. So, a convenient choice in the de Sitter spacetime is the so-called \emph{Bunch-Davies state}, which appears precisely thermal to a free-falling observer in such a space\footnote{ Clearly, this choice is not unique and modifying the vacuum can have consequences on particle production. Our choice is, however, extremely common and widely-used in the literature. See e.g. \cite{vacuum} for an introduction to other possible vacuum choices in inflation.} \cite{Green}. In particular, imposing in our scheme the Bunch-Davies vacuum turns out to be  equivalent to let our solution match the plane wave solution $e^{ik\tau}/\sqrt{2k}$ in the ultraviolet regime $k \gg aH_I$. Thus, we have
\be \label{eq47}
f_k(\tau)= \frac{\sqrt{\pi}}{2}e^{i(\nu +\frac{1}{2})\frac{\pi}{2}} \sqrt{-\tilde{\tau}} H_\nu^{(1)}(-k \tilde{\tau}),
\ee
where $H_\nu^{(1)}$ are first kind Hankel's functions and from Eqs. \eqref{eq46}-\eqref{Hankeq} we get
\be \label{h1index}
\nu^2= \frac{1}{4}+ \frac{2 \Lambda^4}{\mu_2^2 H_I^2}.
\ee


\subsubsection{Scalar field perturbations} \label{sez2A.1}

We assume now that scalar perturbations of the metric are generated by the quantum fluctuations of the inflaton field, as in the standard model of inflation \cite{inflation, referenza4}. We neglect the effects of tensor modes (gravitational waves), which however are expected to produce similar outcomes\footnote{ If the energy-momentum tensor of the inflaton field is diagonal as in our case, it can be shown that also tensor modes are nearly frozen on super-Hubble scales \cite{referenza4}. For what concerns graviton production itself, it has been proven that in this case the relevant terms are those due to the FRW background \cite{referenza2,gravitons}, i.e., perturbative production of gravitons would only give small corrections and thus it is not further investigated here. However, graviton dynamics may be damped due to the creation of scalar particles, as discussed in \cite{referenza2}.} on super-Hubble scales. The most general line element for a perturbed spatially flat FRW universe in case of scalar perturbations reads
\begin{align} \label{scalpert}
ds^2=a^2(\tau) \big[&(1+2\Phi)d\tau^2-2\  \partial_i B\  d\tau dx^i- \notag \\ &\left((1-2\Psi) \delta_{ij}+ D_{ij} E \right) dx^i dx^j \big].
\end{align}
In the longitudinal (or conformal Newtonian) gauge, we set $E=B=0$. For a scalar field, one also obtains\footnote{For the minimally coupled case, this can be shown starting from the non-diagonal part ($i \neq j$) of the (ij)-perturbed Einstein equations. See e.g. \cite{referenza4} for the details. Using a similar argument, one can show that the same result holds during the slow-roll phase of inflation, where the scalar curvature is almost constant.} $\Psi=\Phi$. Accordingly, the perturbation potential $\Psi$ satisfies the differential equation \cite{referenza4, abr}
\be \label{eqpot}
\dot{\Psi}+ \mathcal{H} \Psi=4\pi G \dot{\phi}_0 \delta \phi= \epsilon \mathcal{H}^2 \frac{\delta \phi}{\dot{\phi}_0},
\ee
where we  split the field as
\be \label{totalfield}
\phi=\phi_0(\tau)+\delta \phi({\bf x}, \tau),
\ee
with $\phi_0$ representing the  ``classical'' background field\footnote{Spatial $\phi_0$ expansions are in the form of complex exponentials. Taking infinite wavelengths leads to vanishing momenta, or alternatively to non-oscillations of the field, justifying \emph{de facto} the name  ``classical'' above used, see e.g. \cite{referenza4}.} and $\delta \phi({\bf x}, \tau)$ the quantum fluctuations around $\phi_0$. 
In Eq. \eqref{eqpot}, $G$ is the gravitational constant, $\mathcal{H}=\dot{a}/a$ and $\epsilon$ the usual slow roll parameter\footnote{In our notation, the overdot always refers to derivatives with respect to conformal time, e.g. $\dot{a}= \partial a/\partial \tau$. See Appendix \ref{appA} for the interconnections between conformal, $\tau$, and cosmic time, $t$.}. 
From Eq. \eqref{eq43}, we have
\be \label{eq48}
\mathcal{H}\equiv \frac{\dot{a}}{a}= 
\frac{H_I}{1-H_I\tau},
\ee
with (always) vanishing slow-roll parameter $\epsilon$, given as a consequence of adopting a quasi-de Sitter phase. 
Indubitably, a pure de Sitter spacetime implies $\epsilon=0$ at any time. To overcome this issue, we could easily modify the scale factor by including a slight correction, following the general idea of Ref. \cite{referenza4}. A  plausible modified $a(\tau)$ then reads
\be \label{eq50}
a(\tau)= \frac{1}{(1-H_I\tau)^{1+m}},
\ee
where $m$ is a constant that weakly deviates from $m=0$. For the sake of completeness, in principle we can assume $m=m(\tau)$ instead of a pure constant $m$, in order to properly solve Eq. \eqref{eqpot}. Both the possibilities, namely constant $m$ and $m=m(\tau)$, as anticipated above, are related to the fact that inflation has to be described by a quasi de Sitter phase, first to avoid $\epsilon=0$ and second to guarantee that $a(\tau)$ is an approximate, but suitable, ansatz for our hilltop potential that, only asymptotically, evolves like a de Sitter phase.

With the ansatz \eqref{eq50}, Eq. \eqref{eq48} is slightly modified by
\be \label{eq51}
\mathcal{H}= \frac{(1+m) H_I}{(1-H_I\tau)}
\ee
from which 
\be \label{eq52}
\epsilon= 1- \frac{\dot{\mathcal{H}}}{\mathcal{H}^2}= 
1-\frac{1}{1+m} \simeq m,
\ee
where the last equality on the r.h.s. is true for small $m$ only.
As stated above, by imposing a time-varying $m$ term, the corresponding, more complicated, version of $\epsilon$ would weakly evolve during inflation to guarantee a graceful exit from it. We hereafter leave it fixed throughout the overall inflationary phase only to simplify our calculations and we will require $\epsilon \rightarrow 1$ in order to end inflation.

\subsubsection{Potential solutions at super-Hubble scales}

From now on, we focus on super-Hubble scales, where the condition $k \ll a(\tau)H_I$ holds. This will provide a physically motivated cut-off for the momenta of particles produced, as we will see.  On these scales, it can be shown \cite{referenza4} that $\dot{\phi}_0$ and $\delta \phi$ solve the same equation. The solutions are then related to each other by a function $c({\bf x})$ which depends upon space only:
\be\label{formawww}
\delta \phi= c({\bf x}) \dot{\phi}_0.
\ee
Setting $c({\bf x})= e^{i {\bf q} \cdot {\bf x}}$, we can solve Eq. \eqref{eqpot} for the scale factor \eqref{eq43} and obtain the general form of the potential
\be \label{eq53}
\Psi= \frac{\epsilon H_I-2 c_1(1-H_I\tau)^2}{2(1-H_I\tau)}\ e^{i {\bf q} \cdot {\bf x}}.
\ee
Assuming now that $\Psi(\tau \rightarrow - \infty)=0$, we explicitly get
\be \label{eq54}
\Psi= \frac{\epsilon H_I}{2(1-H_I\tau)}\ e^{i {\bf q} \cdot {\bf x}}. 
\ee
Having a functional form for $\Psi$, we can now compute the perturbation tensor from which our geometric particles arise. 

An interesting point, from Eq. \eqref{eq54} is the following. As $\epsilon$ tends to one, namely as inflation ends, the perturbed potential does not vanish. This is a general feature of inflationary models, not only limited to our choice of $V(\phi)$. Consequently, a suppressing position-dependent exponent in the phase $e^{i {\bf q} \cdot {\bf x}}$ may be requested as cut-off scale in $\Psi$, physically motivated by the fact that once inflation ends the universe is less inhomogeneous than during inflation and, gradually increasing the cosmic scale by cosmic  expansion, one recovers the cosmological principle. 

In other words, a generic form of $\Psi$ for a unspecified potential that violates Eq. \eqref{formawww} may read 
\begin{equation}\label{ik}
\Psi=\mathcal F(\epsilon,\tau,H_I)e^{i {\bf q}(\tau) \cdot {\bf x}}\,,
\end{equation}
with ${\bf q}(\tau)={\bf q}+i{\bf K}\tau$, and ${\bf K}$ a unconstrained positive-definite momentum.

A possible physical motivation to an ansatz of the form \eqref{ik} may lie in the so-called \emph{back-reaction} mechanism, which we now briefly discuss. 


\subsubsection{The issue of back-reaction} \label{sezIIbk}

The perturbed Einstein equations, Eq. \eqref{eqpot}, describe how the inflaton fluctuations affect spacetime geometry during inflation. The next step would be then to compute particle production starting from the perturbation potential $\Psi$, which is the only independent geometric quantity in our framework. However, when particles are produced, they inevitably alter spacetime geometry via their energy-momentum tensor. In other words, particle production induces a back-reaction of the field on the geometry, implying a modification of the original fluctuations $\delta \phi ({\bf x}, \tau)$. 

In Ref. \cite{referenza2}, it has been pointed out that such a mechanism could, in principle, reduce the particle production rate, since any initial inhomogeneity can be damped out as the universe evolution goes on. In our model, we could heuristically overcome this issue by simply changing the instant of time at which particle production is expected to begin, as discussed in Sec. \ref{sez2B.3}. Accordingly, for the moment we neglect the back-reaction mechanism due to its computational complexity, thus preserving the external-field approximation proposed in \cite{referenza2}. Clearly, a self-consistent approach to geometric particle production cannot avoid a proper description of back-reaction, which requires then further investigation. In this direction, a recent gauge-invariant study of back-reaction associated to inflationary particle production has been performed in \cite{altro1}, focusing on a classical approach to cosmological perturbations \cite{backk}.


\subsubsection{Gauge transformations} \label{sez2A.2}

We now need to write the gravitational potential $\Psi$ in the synchronous gauge \footnote{We adopt the synchronous gauge in analogy to seminal papers on geometric particle production \cite{referenza2, cesp}. However, the number of geometric particles produced is a gauge-independent quantity, as discussed in \cite{altro1}, where the more popular longitudinal gauge is also considered.}. In this gauge, the most general scalar perturbation takes the form $h_{ij}=h \delta_{ij}/3+h_{ij}^\parallel$. 

The general procedure to transform from the longitudinal to the synchronous gauge is the following \cite{referenza6}.
Let us consider a general coordinate transformation from a system $x^\mu$ to another $\hat{x}^\mu$
\be \label{S1}
x^\mu \rightarrow \hat{x}^\mu=x^\mu+d^\mu (x^\nu).
\ee
We write the time and the spatial parts separately as
\begin{subequations}
\begin{align} 
&\hat{x}^0= x^0 + \alpha({\bf x},\tau) \label{S2} \\
&\hat{{\bf x}}= {\bf x}+ \nabla \beta({\bf x}, \tau)+ {\boldsymbol \epsilon}({\bf x}, \tau),\ \ \ \ \ \ \ \nabla \cdot {\boldsymbol \epsilon}=0, \label{S3}
\end{align}
\end{subequations}
where the vector $d$ has been divided into a longitudinal component $\nabla \beta$ and a transverse component $\vec{\epsilon}$. 

Let $\hat{x}^\mu$ denote the synchronous coordinates and $x^\mu$ the conformal Newtonian coordinates, with $\hat{x}^\mu=x^\mu+d^\mu$. We have
\begin{subequations}
\begin{align}
& \alpha({\bf x}, \tau)=\dot{\beta}({\bf x}, \tau), \label{S4} \\
& \epsilon_i({\bf x}, \tau)=\epsilon_i({\bf x}), \label{S5} \\
& h_{ij}^\parallel({\bf x}, \tau)=-2 \bigg( \partial_i \partial_j-\frac{1}{3} \delta_{ij} \nabla^2 \bigg) \beta({\bf x}, \tau), \label{S6} \\
& \partial_i \epsilon_j+ \partial_j \epsilon_i=0. \label{S7}
\end{align}
\end{subequations}
and
\begin{subequations}
\begin{align}
    & \psi({\bf x}, \tau)= -\ddot{\beta}({\bf x}, \tau)- \frac{\dot{a}}{a} \dot{\beta}({\bf x}, \tau), \label{S8}\\
    & \phi({\bf x}, \tau)= +\frac{1}{6} h({\bf x}, \tau)+ \frac{1}{3} \nabla^2 \beta({\bf x}, \tau)+ \frac{\dot{a}}{a} \dot{\beta}({\bf x}, \tau) \label{S9}.
\end{align}
\end{subequations}
Now, setting $\Phi=\Psi$, as above stated, and recalling Eq. \eqref{eq54}, we obtain
\be \label{eq55}
\ddot{\beta}({\bf x},\tau)+\frac{\dot{\beta}({\bf x}, \tau)}{1-H_I\tau}H_I+\frac{\epsilon H_I}{2(1-H_I\tau)}\ e^{i {\bf q} \cdot {\bf x}}=0,
\ee
whose general solution is
\be \label{eq56}
\beta({\bf x}, \tau)= \left( -\frac{\epsilon \tau}{2}+ \frac{c_1 \tau^2}{2}+c_2 \right) \ e^{i {\bf q} \cdot {\bf x}}.
\ee

\subsubsection{Perturbation potential}

Let us now focus on the values of the integration constants $c_1$ and $c_2$. Concerning $c_2$ it is easy to see, from Eq. \eqref{S8}, that $\Psi$ vanishes at $\tau\rightarrow0$ independently from the value of $c_2$ that, consequently, is fully-unconstrained. It is straightforward to set $c_2=0$ only to reduce our problem complexity. The situation mostly changes concerning $c_1$. There is no \emph{a priori} reasons to fix it to a given value and apparently $\beta({\bf x},\tau)$ turns out to be quadratic in the conformal time. However, a conceptual caveat suggests how to get it. Indeed, subtracting then Eq. \eqref{S9} from \eqref{S8}, we get
\be \label{eq34}
\ddot{\beta}({\bf x}, \tau)-\frac{2H_I}{1-H\tau} \dot{\beta}({\bf x}, \tau)+ \frac{h({\bf x}, \tau)}{6}=0
\ee
leading to
\be \label{eq35}
h({\bf x}, \tau)= -6 \ddot{\beta}+ \frac{12 H_I}{1-H_I \tau} \dot{\beta}.
\ee
Here, $h({\bf x},\tau)$ would imply non-vanishing perturbations at $-\infty$ that actually diverge, as due to the first-order $\beta$ time-derivative\footnote{Second-order time-derivative implies instead a constant term, almost non-influential for our prescription.}. This fact appears clearly unphysical as we require perturbations to occur during and after inflation, rather than before. Plausibly we are thus forced to set $c_1=0$ to avoid any possible issue. Hence, we get from Eq. \eqref{eq35bis}
\begin{equation}\label{eq35bis}
    h({\bf x}, \tau)= -\frac{6 \epsilon H_I}{1-H_I\tau} e^{i {\bf q} \cdot {\bf x}} = -12 \Psi.
\end{equation}
On super-Hubble scales, the term $h_{ij}^\parallel$ can be neglected. The perturbation tensor in synchronous gauge then reads
\be \label{eq36}
h_{\mu \nu}= \begin{pmatrix} 0 & 0 & 0 & 0\\
0 & 4 \Psi & 0 & 0 \\
0 & 0 & 4 \Psi &0 \\
0 & 0 & 0 & 4 \Psi
\end{pmatrix},
\ee
from which the line element 
\be \label{eq37}
ds^2=a^2(\tau)\left[d\tau^2-\delta_{ij}(1-4\Psi)dx^i dx^j  \right].
\ee
We are now ready to compute the corresponding geometric particle production.


\subsection{Geometric particle production} \label{sez2B.3}

In the external-field approximation, we can describe the interaction of the inflaton with spacetime geometry at first perturbative order via the Lagrangian \cite{referenza2}
\be \label{eqlag}
\mathcal{L}_I= -\frac{1}{2} \sqrt{-g_{(0)}} H^{\mu \nu}T_{\mu \nu}^{(0)}\,,
\ee
where $g_{\mu \nu}^{(0)}\equiv a^2(\tau) \eta_{\mu \nu}$, $H_{\mu \nu}=a^2(\tau)h_{\mu \nu}$ and $T_{\mu \nu}^{(0)}$ is the zero-order energy-momentum tensor, namely
\begin{align} \label{emtens}
T_{\mu \nu}^{(0)}= &\partial_\mu \phi \partial_\nu \phi- \frac{1}{2}g_{\mu \nu}^{(0)} \left[ g^{\rho \sigma}_{(0)} \partial_\rho \phi \partial_\sigma \phi-2\Lambda^4(1-\phi^2/\mu_2^2) \right]\notag \\
&-\frac{1}{6}\left[\nabla_\mu \partial_\nu-g_{\mu \nu}^{(0)}\nabla^\rho \nabla_\rho+R_{\mu \nu}^{(0)}-\frac{1}{2} R^{(0)}g_{\mu \nu}^{(0)} \right]\phi^2.
\end{align}
The first-order $\hat S$-matrix can be obtained by Dyson's expansion formula (see e.g. \cite{lanca})
\begin{equation}\label{Dyson}
\hat S=\hat T\exp^{-i\int_{-\infty}^{+\infty}d^4x \mathcal H_I},
\end{equation}
where $\mathcal{H}_I$ is the Hamiltonian density in interacting picture and $\hat T$ the time-ordering operator. 

The exponential form of Dyson’s expansion is not practical, since the integral in the exponent cannot be computed exactly. We may then expand out Eq. \eqref{Dyson} at first order, recalling that the interaction Hamiltonian is smaller than the background one. As  $ \mathcal{H}_I=-\mathcal{L}_I$ in our model \cite{lee, abers}, following the standard procedure in Dyson's expansion we get
\be \label{smatr}
\hat S \simeq 1+i \hat{T} \int d^4x\ \mathcal{L}_I.
\ee
Accordingly, the second order particle number density at time $\tau^*$ is
\be \label{ndens}
N^{(2)}(\tau^*)= \frac{a^{-3}(\tau^*)}{\left(2\pi  \right)^{3}} \int d^3k\  d^3p\  \lvert \langle 0 \lvert \hat S \rvert k,p \rangle \rvert^2. \\[5pt]
\ee
We remark that second order terms are not required in the $\hat S$-matrix expansion \eqref{smatr}, since the interaction Lagrangian is still quadratic in the field at second geometric order, thus contributing at higher orders to the particle number density. Moreover, in Eq. \eqref{ndens} we have assumed that no ``quantum" particle production is involved, namely the Bogoliubov coefficients $\beta_k$ and $\beta_p$ obtained in \cite{referenza2} have been neglected. In Appendix \ref{appB} we discuss the generalization of Eq. \eqref{ndens} to the case of non-zero Bogoliubov coefficients. Quantum particle production is also responsible for the generation of particle-antiparticle pairs at zero and first geometric order \cite{par, referenza2, cesp}, as we will discuss in Sec. \ref{sez3}.

Coming back to Eq. \eqref{ndens}, the probability amplitude for particle pair creation can be derived from Eq. \eqref{eqlag}-\eqref{smatr}, namely
\begin{widetext}
\begin{align} \label{eq42}
\langle 0 \lvert \hat S \rvert k,p \rangle=& -\frac{i}{2} \int d^4x\ 2a^4  H^{ij} \bigg[ \partial_{(i} \phi_k^* \partial_{j)}\phi_p^*-\frac{1}{2}\eta_{ij} \eta^{ab} \partial_{(a} \phi_k^* \partial_{b)}\phi_p^*+g_{ij}^{(0)}\Lambda^4\left(-\frac{\phi^*_k \phi^*_p}{\mu_2^2}\right)  \notag \\
&\ \ \ \ \ \ \ \ \ \ \ \ \ \ \ \ \ \ \ \ \ \ \ \ -\frac{1}{6} \left( \nabla_i \partial_j-g_{ij}^{(0)}\nabla^a \nabla_a+R_{ij}^{(0)}-\frac{1}{2} R^{(0)}g_{ij}^{(0)} \right)\phi_k^* \phi_p^* \bigg]\,, 
\end{align}
\end{widetext}
with $i,j=1,2,3$ as consequence of working in the synchronous gauge. We have also defined the field modes 
\be \label{modes}
\phi_k ({\bf x}, \tau)= \frac{f_k(\tau)}{(2 \pi)^{3/2} a(\tau)} e^{i {\bf k}\cdot {\bf x}},
\ee
which can be derived from Eqs. \eqref{eq9} together with the solutions Eqs. \eqref{eq47}. 

On super-Hubble scales, these modes can be written as \cite{referenza4}
\begin{widetext}
\be \label{eq64}
\phi_k= \frac{1}{(2 \pi)^{3/2}} e^{i\left(\nu-\frac{1}{2}\right)\frac{\pi}{2}}  2^{\nu-\frac{3}{2}} \frac{\Gamma(\nu)}{\Gamma(3/2)} \frac{H_I}{\sqrt{2k^3}} \left(\frac{k}{aH_I} \right)^{\frac{3}{2}-\nu} e^{i {\bf k}\cdot {\bf x}}.
\ee
\end{widetext}
Exploiting now the fact that the perturbation tensor is diagonal and writing explicitly all the curvatures, Eq. \eqref{eq42} can be recast in the compact form
\be \label{eq57}
\langle 0 \lvert \hat S \rvert k,p \rangle= -\frac{i}{2} \int d^4x\  2a^4\left(A_1({\bf x},\tau)+A_2({\bf x},\tau)+A_3({\bf x},\tau) \right),
\ee
where $A_i({\bf x}, \tau)$ are the only non-zero contributions to the probability amplitude, namely
\begin{widetext}
\begin{align} \label{eq65}
    A_i({\bf x},\tau)=H^{ii}\cdot \bigg[&\partial_i \phi_k^* \partial_i \phi_p^*+\frac{1}{2} \eta^{ab} \partial_a \phi_k^* \partial_b \phi_p^*+a^2 \Lambda^4 \left( \frac{\phi_k^* \phi_p^*}{\mu_2^2}\right) -\frac{1}{6} \left(\partial_i \partial_i+\frac{\dot{a}}{a}\partial_0+\eta^{ab}\partial_a \partial_b-\left(\frac{\ddot{a}}{a}+\left( \frac{\dot{a}}{a} \right)^2 \right)+3\frac{\ddot{a}}{a}  \right) \phi_k^* \phi_p^* \bigg].\notag\\
   \end{align}
\end{widetext}


\subsubsection{Dark matter from ``geometric particles''?} \label{sez2B.4}

With all the above ingredients, we can now compute the final number density of geometric particles produced, namely $N^{(2)}(\tau)$ at $\tau=0$. As anticipated, these are interpreted in terms of dark matter quasiparticles. Dark matter seems the most plausible candidate in our model, since it only interacts gravitationally with ordinary matter and, in fact, the way of obtaining it derives from the Yukawa-like potential only. We expect that any particle pair creation, got from purely quantum processes, becomes subdominant over quasiparticles obtained directly from vacuum fluctuations  \cite{Lythp}, as above discussed. 

Hence, to determine dark matter microphysics and properties, we first need to specify initial inflationary settings, \textit{i.e.}, to properly define super-Hubble scales, introducing a cut-off scale to have enough e-foldings, say $N$, that are needful to speed up the universe during inflation \cite{inflation}, having
\be \label{eq70}
N= \int\ dt H(t)=  \int\ d\tau\ \frac{H_I}{1-H_I\tau} \simeq 60,
\ee
where conventionally we took $60$ as minimal number of e-foldings. We thus obtain  $\log(1-H_I \tau) \bigg \lvert_0^{t_I}= 60$,
where $t_I<0$ is assumed to be the initial time for inflation, and it can be inferred once the fixed values are imposed on our free parameters.

Lying on super-Hubble scales, namely 
\be \label{eq71}
\frac{k}{a(\tau)H_I} \ll 1 \implies k \ll a(\tau)H_I,
\ee
quantum fluctuations of the inflaton field become classical\footnote{The same expression of Eq. \eqref{eq71} formally holds for $p$ also.}, \textit{i.e.}, they no longer oscillate in time (cfr. Eq. \eqref{eq64}). In this respect, we can properly get  particles only after horizon exit\footnote{A more detailed discussion on the notion of particle at horizon crossing can be found in chapter 24 of Ref. \cite{Lythp}.}.

Easily, Eq. \eqref{eq64} is valid throughout all the inflationary epoch, as we take the minimum of $aH_I$, say $a(\tau_I)H_I$, as required cut-off. This ensures that the field modes are described by Eq. \eqref{eq64} as $\tau > \tau_I$. However, since $a(\tau_I) \propto \exp(-60)$, this choice would result in a very small cut-off for particle momenta. Consequently, from a genuine physical perspective, this issue is healed  by assuming that geometric particle production started at $t_i > t_I$, \textit{i.e.}, not exactly at the beginning of the inflationary era.  In our computation we can show that, in view of our effective potential parameters, realistic values for the dark matter number density may be obtained within the range $t_i \in [-\exp(45)/H_I, -\exp(40)/H_I]$. For completeness, however, we remark that our choice of time-independent cut-off inevitably leads to \emph{underestimating the total number density}. This happens because we essentially neglect all the momenta whose horizon crossing is subsequent the time $\tau_i$. 

\subsubsection{Constraints on the effective potential}

Concerning the requirements of our effective potential, we invoke the following basic demands. 

\begin{itemize}
    \item[--] Since inflation is thought to follow a quantum gravity regime, we  expect vacuum energy scales to lie on Planck mass scales, namely
    \be \label{enegyscale}
    \Lambda^4 \simeq M_{\rm pl}^4,
    \ee
    where $M_{\text{pl}}=1.22 \times 10^{19}$ GeV is the Planck mass. This ansatz agrees with current understanding about the value of the cosmological constant as predicted by quantum field theory fluctuations \cite{wein1}.
    \item[--] The corresponding slightly evolving Hubble rate during inflation is therefore 
    \be \label{eq74}
    H^2(t)\equiv H^2_I\simeq \frac{8 \pi G}{3} \Lambda^4,
    \ee 
    and Planck satellite data \cite{referenza3} impose the following constraint (at a 95 \% confidence level):
    \be \label{eq76}
    \frac{H_I}{M_{\text{pl}}} < 2.5 \times 10^{-5},
    \ee
    which accordingly would give $\Lambda^4 \lesssim 10^{65}$ GeV$^4$. In particular, this energy scale for vacuum energy is the typical regime of spontaneous symmetry breaking in grand unified theories \cite{referenza10, coles}.
    \item[--] The minimally coupled hilltop quadratic potential requires  \cite{referenza3}
    \be \label{eq77}
    0.3 < \log_{10} (\mu_2/M_{pl}) < 4.85,
    \ee
    namely $2\ M_{\rm pl} \lesssim \mu_2 \lesssim 10^5\ M_{\rm pl}$. Our effective potential, instead,  includes  additional field-curvature coupling contribution, that provide relevant consequences on inflation. Nevertheless, in case of conformal coupling, large  $\Lambda^4$ values would result in vacuum energy domination. Hence, it appears licit to consider the prescription of Eq. \eqref{eq77} in our computation as prior for Eq. \eqref{effectivepotential}.
\end{itemize}
Concerning the choice of the slow-roll parameter, we have previously discussed  that small deviations from a pure de Sitter evolution are required in order to have a non-zero $\epsilon$. Since we are dealing with inhomogeneities at a perturbative level, we also have to satisfy \cite{referenza2, cesp}
\be \label{tenspert}
\lvert h_{ij}(x) \rvert \ll 1.
\ee  
Hence, by virtue of Eq. \eqref{eq54} we see that in order to preserve the perturbative treatment, we further need $\epsilon H_I \ll 1$. 
\begin{figure}
    \centering
    \includegraphics[scale=0.68]{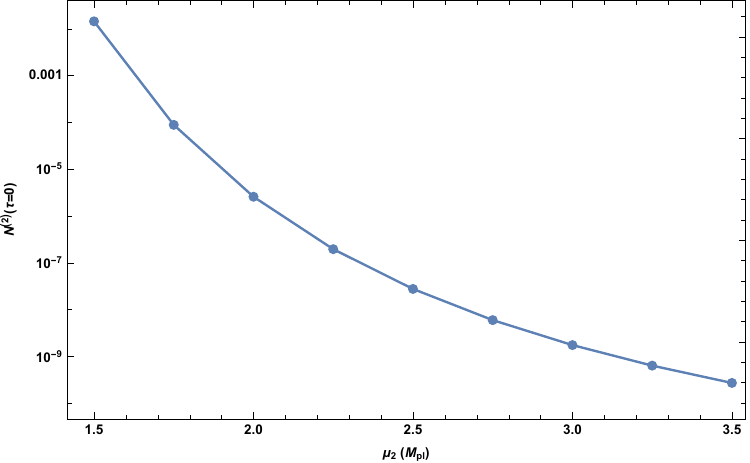}
    \caption{Number density $N^{(2)}$ in GeV$^3$ as function of the hilltop width, $\mu_2$, in units of $M_{\rm pl}$. The other involved parameters are: $\Lambda^4=10^{64}$ GeV$^4$, $\epsilon=10^{-14}$, $\tau_i=-\exp(40)/H_I$. The value $40$ is conventionally chosen inside the whole interval in which inflation occurs, as explained in the text.}
    \label{fig1}
\end{figure}
In this respect, we draw in Fig. \ref{fig1} the number density of geometric particles, namely $N^{(2)}(0)$, for given values of the hilltop parameter $\mu_2$. In Fig. \ref{fig2} we show the dependence of the number density on the vacuum energy term driving inflation, $\Lambda^4$.
\begin{figure}
    \centering
    \includegraphics[scale=0.68]{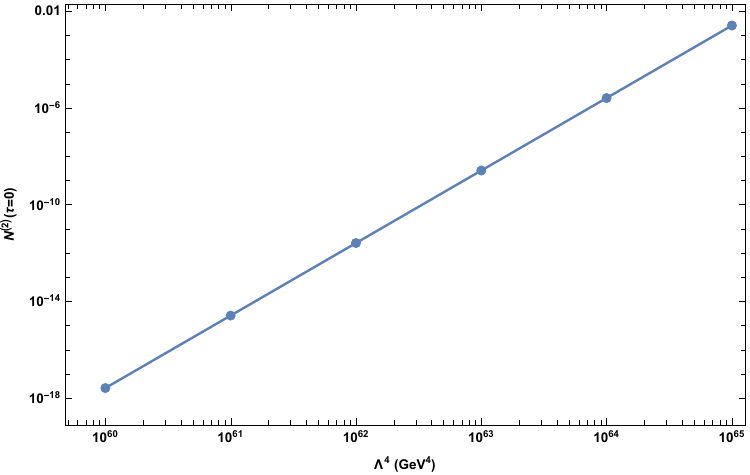}
    \caption{Number density $N^{(2)}$ in GeV$^3$ as function of vacuum energy. We have set $\mu_2=2\ M_{\rm pl}$, $\tau_i=-\exp(40)/H_I$ and $\epsilon$ is chosen so that $\epsilon H_I= {\rm const}$. The value $40$ is conventionally chosen inside the whole interval in which inflation occurs, as explained in the text.}
    \label{fig2}
\end{figure}

\subsubsection{Cut-off scales and vacuum energy}

A further inspection of Eqs. \eqref{h1index} and \eqref{eq64} reveals that the modes of the field become exactly ``frozen" on super-Hubble scales if $\nu \simeq 3/2$, namely
\be \label{froz}
\mu_2 \simeq 0.42 \times 10^{19}\ {\rm GeV}= 0.34\ M_{\rm pl}.
\ee
This value is well outside the range provided in Eq. \eqref{eq77} that, however, we know to be valid only in case of minimally-coupled inflaton. Smaller $\mu_2$ values generally lead to a larger number density. Accordingly, if we require Eq. \eqref{froz} to hold in general, then two possibilities arise:

\begin{itemize}
    \item[--] the cut-off scale on particle momenta is forced to be much smaller, in order to preserve realistic values for the number density $N^{(2)}(0)$;
    \item[--] alternatively, vacuum energy should be several orders of magnitude below Planck energy scale, \textit{i.e.}, closer to the scales of the particle physics standard model.
\end{itemize}

The latter possibility is still an open and promising scenario. Constraining $N^{(2)}(0)$ requires to know the cut-off scale that \emph{a priori} cannot be known. Disclosing  how to constrain the vacuum energy cut-off scales  deserves further investigation and will be object of future works. 



\subsection{Phase B: Exiting inflation} \label{sez2C}

We featured the inflationary epoch by using the de Sitter solution of Eq. \eqref{eq43}. This represented a suitable approximation that, however, fails to be predictive during the reheating transition, \textit{i.e.}, as inflation ends. In particular, at the end of inflation Eq. \eqref{eq74} no longer holds and the behavior of the scale factor is not determined by the sole vacuum energy, $\Lambda^4$. Instead, it depends on the full effective potential $V^{\text{eff}}$ that, consequently, should be evaluated \emph{in toto}. 

In particular, the potential $V^{\text{eff}}$ of Eq. \eqref{effectivepotential} also includes the coupling to the Ricci scalar curvature, which is crucial to interpret geometric particles as dark matter quasiparticles. However, this interacting term grows as $\phi$ increases. By construction, it grows when the hilltop potential evolves towards its minimum. Accordingly, the presence of such coupling would not allow a graceful exit from inflation, since the full potential never reaches its minimum. In principle, this may suggest that a more complicated version the single-field inflationary potential would be required in order to properly address the transition from inflation to reheating. For instance, in \cite{DLM} a Morse potential reducing to the Starobinski one is investigating, unifying \emph{de facto} inflation with dark energy.

\subsubsection{Approximating the reheating phase}

Here, as naive estimation we can assume that during reheating the background geometry behaves \emph{in average} as a matter dominated universe \cite{referenza7, referenza11} and so, accordingly to this hypothesis, we select a scale factor that fulfills an Einstein-de Sitter (EdS) universe dominated by matter through 
\be \label{scalereh}
a(\tau)= \left( 1+ \frac{H_I \tau}{2} \right)^2,\ \ \ \ \ 0 < \tau < \tau_{r}
\ee
where we denote with $\tau_{r}$ the time at which reheating is expected to end.

We now need continuity of each epoch, without passing through any transition. To do so, we notice that Eq. \eqref{scalereh} ensures the validity of the matching conditions \cite{referenza12} at time $\tau=0$, essentially implying the continuity of the scale factor and Hubble parameter on the junction hypersurface.  For $\tau >0$, the zero-order scalar curvature takes the form:
\be \label{riccireh}
R^{(0)}= \frac{3 H_I^2}{(1+\frac{H_I\tau}{2})^6},\ \ \ \ \ 0 < \tau < \tau_r
\ee
and introducing the usual slow-roll parameter
\be
\epsilon= \frac{1}{16 \pi G}\left(\frac{V^\prime}{V} \right)^2, \label{slowroll}\\
\ee
we can exploit Eq. \eqref{effectivepotential} and set $\epsilon=1$ to obtain a realistic value $\phi_{\rm end}$ for the field at the end of inflation. For $R^{(0)}(\tau >10^{-12})$ the hilltop contribution dominates over the field-curvature coupling in $V^{\rm eff}$, and  we expect a result close to the one obtained for minimally coupled hilltop models  \cite{referenza13}, namely
\be \label{exit}
2q^2 \frac{(\phi_{\rm end}/\mu_2)^2}{(1-(\phi_{\rm end}/\mu_2)^2)^2}=1\,,
\ee
having defined $q \equiv M_{\rm pl}/\mu_2$. We obtain the following expansion for small $q$,
\be \label{hilexp}
\phi_{\rm end}= \mu_2\left(1-\frac{1}{\sqrt{2}}q + \frac{1}{4}q^2+ \mathcal{O}(q^3) \right)\,,
\ee
so that we have a nonzero remaining contribution to the hilltop component of the potential.
We expect this remaining contribution to the potential to be responsible for baryonic particle creation, as usually discussed in preheating and reheating models (see e.g. \cite{referenza11}).

\subsubsection{Approximating the radiation dominated phase}

As the reheating stops, the subsequent phase of radiation domination can be modeled by a scale factor of a EdS universe of the form
\be \label{radiation}
a(\tau)=b\tau+c,\ \ \ \ \ \tau_{r} < \tau < \tau_m
\ee
where the constants $b$ and $c$ are determined by again imposing the matching conditions on the matching hypersurface from reheating to radiation phase, at time $\tau=\tau_r$. The quantity $\tau_m$ denotes the instant of time at which transition to the matter-dominated era is expected to happen.
During radiation domination, the corresponding EdS Hubble parameter and temperature satisfy  \cite{referenza10}
\begin{align}\label{hrad}
H &\sim \sqrt{G}\,T^2,
\end{align}
where $T$ is the corresponding temperature of the universe.  The expected dark matter energy density at $\tau_r$ is then\footnote{For the sake of completeness, the subsequent dark matter contribution \emph{is not} pressureless as in the standard cosmological model, but has a non-negligible term that agrees with the one found in Ref. \cite{LM}. This term, however, is absolutely negligible at the reheating time and does not affect the universe dynamics. We will show later in the text that its magnitude can be associated to current observations of the cosmological constant, tackling the coincidence problem.}
\begin{align}\label{dmdensity}
\rho^{\rm DM}\left (\tau_{r}\right) &\simeq  (1+z_{r})^3 \rho^{\rm DM}_0, 
\end{align} 
where $\rho_0^{\rm DM}$ is the current value got at redshift $z=0$, namely  $\rho^{\rm DM}_0\simeq 0.25\ \rho_{\rm cr}$, with $\rho_{\rm cr}\equiv 3H_0^2/ 8\pi G$ and $H_0$ is the Hubble constant. Further, we introduced the redshift $z_r$ that certifies the beginning of the radiation phase. Since we are dealing with the radiation-dominated epoch, $z_r$ can be obtained within a EdS universe dominated by radiation only, \textit{i.e.},
\be
H(z) \simeq H_0^2 \Omega_{0r} (1+z)^4.
\ee
Here, $\Omega_{0r}$ is the today radiation density, say $\Omega_{0r}\equiv \rho_0^r/\rho_{\rm cr} \simeq 9.29 \times 10^{-5}$ \cite{planck13}.
Using now the ansatz made in Eq. \eqref{hrad}, we get
\be \label{redsh}
z_{r}\equiv z\left(T_{r}\right) \simeq \sqrt[4]{\frac{G T^4_{r} }{H_0^2\  \Omega_{0r}}},
\ee
where $T_{r}$ is the temperature corresponding to $z_r$.

\section{Dark matter  constituent} \label{sez2E}

Assuming the whole dark matter is produced during inflation, via the geometric mechanism described in Sec. \ref{sez2B.3}, we could in principle estimate its mass. By construction we have $\rho^{\rm DM}= m^* N^{(2)}$, where $m^*$ is the mass of the dark matter candidate. Thus, by virtue of Eqs. \eqref{dmdensity} and \eqref{redsh}, we easily get
\be \label{dmdensity2}
m^*(T_{r})=\frac{\rho^{\rm DM}}{N^{(2)}} \simeq \left(\frac{G T_{r}^4}{H_0^2\  \Omega_{0r}}\right)^{3/4} \frac{\rho^{\rm DM}_0}{N^{(2)}},
\ee
and both densities might be computed at $\tau=\tau_r$. However, since this time is \emph{a priori} unknown, we cannot use Eq. \eqref{scalereh} to compute the  normalization factor for $N^{(2)}(\tau_r)$. This issue may be healed assuming, for instance, that $\tau_r$ is  small enough to show $a(\tau=\tau_r) \simeq a(\tau=0)=1$. We therefore simply follow the latter approach, just noticing that any  larger  $\tau_r$ would only slightly modify the normalization factor for $N^{(2)}$, as confirmed in Eq. \eqref{ndens}. 

\begin{figure}
    \centering
    \includegraphics[scale=0.68]{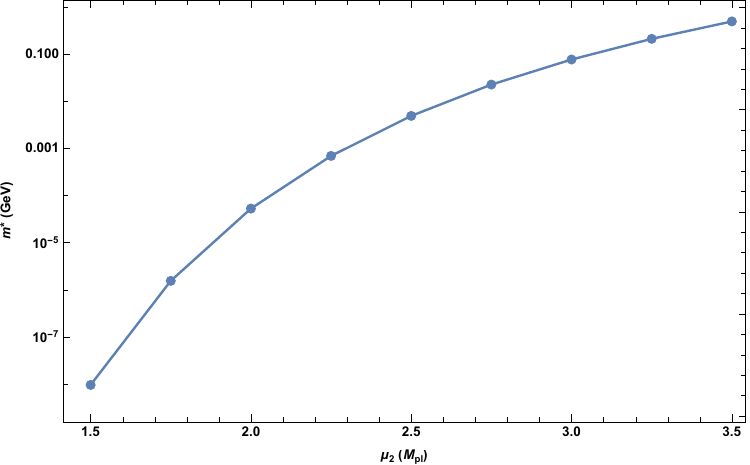}
    \caption{Mass of the dark matter candidate (GeV) as function of the hilltop parameter $\mu_2$. The other parameters are: $\Lambda^4=10^{64}$ GeV$^4$, $\epsilon=10^{-14}$, $\tau_i=-\exp(40)/H_I$ and $T_r=1$ GeV.}
    \label{fig3}
\end{figure}

\begin{figure}
    \centering
    \includegraphics[scale=0.68]{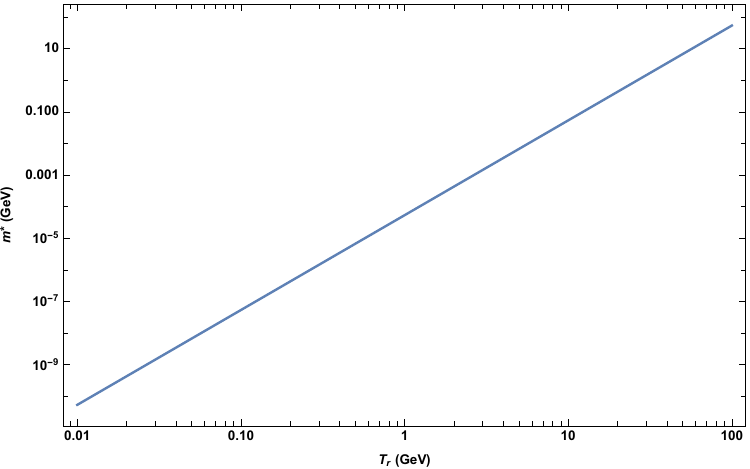}
    \caption{Mass of the dark matter candidate in GeV as function of the temperature $T_r$ at the beginning of radiation phase. The other employed parameters are: $\Lambda^4=10^{64}$ GeV$^4$, $\epsilon=10^{-14}$, $\tau_i=-\exp(40)/H_I$ and $\mu_2=2\ M_{\rm pl}$.}
    \label{figT}
\end{figure}

Hence, fixing the temperature $T_{r}$ and employing the parameters $\Lambda^4\simeq 10^{64}$ GeV$^4$, $\epsilon=10^{-14}$ introduced in Fig. \ref{fig1}, we can compute the value of the mass $m^*$ for $\mu_2$ in a given interval, as reported in Tab. \ref{tab1} and prompted in Fig. \ref{fig3}. In Tab. \ref{tab1} we also show that a larger (in absolute value) $\tau_i$ would lead to larger values for the mass of the dark matter candidate. As already discussed, this is due to the fact that larger $\lvert \tau_i \rvert$ would result in smaller values for the momentum cut-off and thus less particles produced. In Fig. \ref{figT} we show the dependence of the mass $m^*$ on the temperature $T_{r}$, for $\mu_2=2 M_{\rm pl}$.

\begin{table}[h!] 
\centering
 \begin{tabular}{||c c c ||} 
 
 \hline
 $\tau_i$ (GeV$^{-1}$) & $\mu_2 (M_{\rm pl})$ & $m^* (\text{GeV})$ \ \\ [0.5ex] 
 \hline\hline
 $\,$ & $\,$ & $\,$ \\
$-\frac{\exp(40)}{H_I}$ & $1.5$ & $9.45 \times 10^{-10}$ \\ 
 & $2.0$ & $5.29 \times 10^{-5}$ \\
 & $2.5$ & $4.94 \times 10^{-3}$  \\
 & $3.0$ & $7.80 \times 10^{-2}$ \\
 & $3.5$ & $0.503$ \\[3 pt]
$-\frac{\exp(45)}{H_I}$ & $1.5$ & $0.661$ \\ 
 & $2.0$ & $8.31 \times 10^{3}$ \\
 & $2.5$ & $1.15 \times 10^6$  \\
 & $3.0$ & $2.26 \times 10^{7}$ \\
 & $3.5$ & $1.66 \times 10^{8}$ \\ [1ex] 
 \hline
 \end{tabular}
 \caption{Table of masses $m^*$ of the geometric dark matter candidate for given values of $\tau_i$ and the hilltop parameter $\mu_2$, assuming conventionally $T_{r}=1$ GeV. The numbers $40$ and $45$ are arbitrarily chosen to reduce the interval in which dark matter is produced, as explained in detail in the text. } 
 \label{tab1}
\end{table}

We notice then that the total amount of dark matter present in the universe could in principle be traced back to a geometric particle production mechanism. We remark again that our results critically depend on the momentum cut-off scales, introduced in Sec. \ref{sez2B.4}, which is intimately related to the initial ansatz for the scale factor during inflation. A larger cut-off would result in a larger number of particles produced and, therefore, smaller values for the mass $m^*$. We also underline that the initial temperature of the radiation phase is in principle a model dependent quantity. \\
\begin{figure} [ht!]
    \centering
    \includegraphics[scale=0.68]{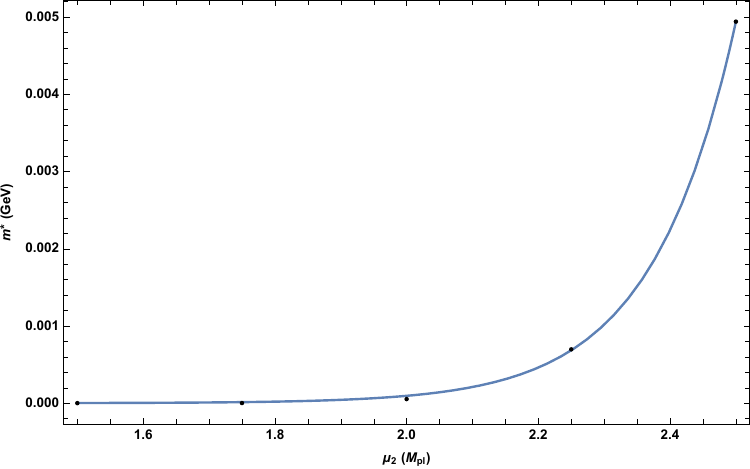}
    \caption{Same points of Fig. \ref{fig3} (black dots) in the range $\mu_2 \in [1.5,2.5]\, M_{\rm pl}$ fitted with a test function found under the form of an exponential: $f(\mu_2)=a\,\exp(b\mu_2+c)$. The fit has been carried out by the \emph{FindFit} command in \textit{Wolfram Mathematica}. The best fit values are found as: $a=4.969 \times 10^{-11}$, $b=7.907$, $c=-1.353$.}
    \label{fig4}
\end{figure}
In Fig. \ref{fig4} we show that the points of Tab. \ref{tab1} fit well with an exponential function, provided $\mu_2$ remains close to the lower bound imposed by Planck (cfr. Eq. \eqref{eq77}).


\section{Coincidence and fine-tuning problems} \label{sez2F}

The above-developed strategy is proposing a  toy model mechanism that transforms  vacuum energy into particles. Since dark matter here arises from the coupling between inflaton and curvature, at a perturbative level, we baptized it as due to geometric particles, whose collective behavior turns out to be stable throughout the universe evolution, having therefore  quasi-particle constituents, as argued in Ref. \cite{BLM}. 

In view of this, we here focus on the universe dynamics and we show how to obtain a heuristic argument to alleviate the coincidence and fine-tuning problems plaguing the standard cosmological background model. To do so, since we have assumed continuity between  the inhomogeneous and homogeneous epochs, \textit{i.e.}, the cosmic dynamics is smooth and no discontinuities are expected, we can proceed as schematically listed below.

\begin{itemize}
    \item[--] We ask that, in addition to continuity between epochs, the Israel--Darmois junction conditions  hold \cite{referenza12}. These conditions require that, on the spacelike hypersurface representing the junction time, the two metric tensors induced by each universe coincide, as well as the two extrinsic curvatures 
    (see also \cite[\S 21.13]{MTW}). 
    \item[--] For each metric, namely for the homogeneous and inhomogeneous spacetime, we evaluate the corresponding  energy and pressure. We call them  $\rho_{1;2}$ and  $P_{1;2}$, where conventionally we refer to subscripts $1;2$ as inhomogeneous and homogeneous metrics, respectively.
    \item[--] Since the overall energy is conserved by construction, we calculate the \emph{pressure jump}, \textit{i.e.}, the difference $P^{(1)}-P^{(2)}$. If the latter would be proportional to the critical density of the universe today, or smaller, then the energy transformed into geometric particles leaves the pressure magnitude today of the same order of current observations. 
\end{itemize}


\subsection{The role of the bare cosmological constant} \label{sezIVA}

The last item, essential for our purposes, occurs because at the end of inflation the energy of the initial inhomogeneous universe is much smaller than vacuum energy. The energy lost, and transformed into geometric particles, is responsible for $\rho^{(1)}$ magnitude. This fact can naturally fix the fine-tuning issue today. Indeed, if $\rho^{(2)}$ is equal to the density at the end of inflation, \textit{i.e.}, without the degrees of freedom of quantum vacuum energy, the fine-tuning issue is not a well-posed problem, but rather it turns out to be  naturally overcome. Accordingly, since $\rho^{(2)}=\rho^{(1)}$  if we get $P^{(2)}-P^{(1)}\leq \rho_{cr}$ then the two magnitudes of pressure would be comparable and so $P^{(2)}$, \textit{i.e.}, the pressure today, will be proportional to the matter density at late-times alleviating the coincidence problem \cite{whynow}.

Following the nomenclature of Sec. \ref{introduzione},  we can schematically sketch the corresponding net values reached by  $\Lambda$ as:
\begin{equation}
    \Lambda\simeq\begin{cases}
        \rho_{vac},\quad &{\rm Before\,\, inflation,} \\
        \Lambda_B,\quad &{\rm After\,\, inflation.}
    \end{cases}
\end{equation}
In this picture,  $\Lambda_B$ is the magnitude inferred once we evaluate $P^{(1)}-P^{(2)}$. Thus, the bare cosmological constant arises since not all the vacuum energy is cancelled. In fact, the pressure jump, namely $P^{(1)}-P^{(2)}$ is due to the fact that vacuum energy is not completely fine-tuned to give geometric particles, but rather a (large) fraction of it provides particles. Hence, if we indicate with $\Lambda_{geom}\leq \rho_{vac}$ the amount used for getting particles, we conclude
\be \label{geocanc}
\Lambda_B=\rho_{vac}-\Lambda_{geom}.
\ee 
In the above equation, we neglected the fact that, at the end of inflation, there is also a remaining contribution due to the inflaton potential, since $\phi_{\rm end}/\mu_2$ may have small deviations from unity, as reported in Eq. \eqref{hilexp}. However, as already explained, we expect this contribution to be responsible for ordinary baryonic matter production during reheating and, for this reason, it is not involved in our argument here.

Eq. \eqref{geocanc} is then true if inflation ends before cancelling completely the overall vacuum energy pressure, leaving a residual constant pressure to contribute the spatial part of the energy momentum tensor \emph{after inflation}, being proportional to $\Lambda_B$. In such a picture, $\Lambda_B$ is therefore reinterpreted as the difference of pressures before and after the transition that is associated to the particle production. This mechanism fully-agrees with the one presented in Ref. \cite{LM}, but the here-adopted hilltop potential differs from the one prompted in \cite{DLM}.

It is finally useful to remark that we use the end of inflation, assuming that $\epsilon\rightarrow1$, to compute the contribution of $\Lambda_{geom}$. This furnishes the remaining contribution to the potential that leads to baryonic particle creation. The corresponding value for $\Lambda_{geom}$ is therefore not fine-tuned but determined by when the inflationary time ends. 


\subsection{The  junction conditions} \label{sezIVB}

We now consider the transition from the inhomogeneous inflationary scenario to the matter-dominated reheating previously discussed. We write for the two spacetimes the corresponding line elements to hold\footnote{We here report both the spacetimes in conformal time representations. This choice is arbitrary: analogous results can be found  using cosmic time instead of $\tau_1$ and $\tau_2$.}
\begin{subequations}
\begin{align}\label{eq37like}
g_1&=a_1^2(\tau_1)\left[d\tau_1^2-\delta_{ij}\Gamma^2\,dx^i dx^j  \right],\\
g_2&=a_2^2(\tau_2)\left[d\tau_2^2-\delta_{ij}dX^i dX^j  \right],\label{eq:eq37like-b}
\end{align} 
\end{subequations}
with $\Gamma\equiv \sqrt{1-4\Psi}$ and \emph{a priori} $\tau_1\neq\tau_2$, $x_i\neq X_i$. Here the perturbed metric $g_1$ is associated to Eq. \eqref{eq37} whereas $g_2$ is the current spatially-flat homogeneous and isotropic FRW spacetime, still valid up to our time by simply fulfilling the cosmological principle. We evaluate the Israel--Darmois junction conditions and we assume the following recipe:

\begin{itemize}
    \item[--] we measure time  regardless the cosmological epoch, leading to  $\tau_1=\tau_2$;
    \item[--] the angular part of both the spacetimes remains unaltered before and after the matching.
\end{itemize}

The equivalence of the  metric tensors \eqref{eq37like}--\eqref{eq:eq37like-b}  induced on $\tau=0$ (chosen as the junction time between the two phases) gives that both spatial line elements  must coincide, $x_i=X_j$, up to radial rescaling. This in turn implies a condition that the metric coefficient $\Gamma$ containing the potential $\Psi$ must satisfy on the matching hypersurface $\tau=0$: 
\begin{equation}\label{eq:match1}
    \Gamma=\frac{a_2(0)}{a_1(0)}.
\end{equation}
Moreover, equivalence of the two extrinsic curvatures gives the further condition 
\begin{equation}\label{eq:match2}
\Gamma_{,\tau}=H_2(0)-H_1(0),
\end{equation}
where $\Gamma_{,\tau}\equiv\frac{\partial\Gamma}{\partial\tau}$, evaluated at $\tau=0$ again.

When we discussed about reheating time, we required matching continuity of our functions. So, in analogy, assuming the universe not to pass through any transition and/or discontinuity, we take its size and radius to be continuous. Consequently, from \eqref{eq:match1}--\eqref{eq:match2}, it is licit to write down
\begin{subequations}
\begin{align}
 a_2(0)&=a_1(0)\,,\label{disuman1}\\
H_2(0)&=H_1(0)\,.\label{disuman2}
\end{align}
\end{subequations}
From Eqs. \eqref{eq:match1}--\eqref{eq:match2}, by virtue of the above relations, we get the intriguing fact that $\Psi$ must be constant on the junction hypersurface, in order to  permit the matching between the two spacetimes to occur.

By construction from Einstein's equations, one expects the pressure term to be  proportional to the second derivative of $\Gamma$ with respect to $\tau$, namely $\Gamma_{,\tau\tau}$. If the pressure difference between the first and second stage of our spacetimes is proportional to $\rho_{cr}$, then the coincidence problem would be alleviated. 

Bearing in mind Eqs. \eqref{eq:match1} and \eqref{eq:match2} with the recipe of  Eqs.  \eqref{disuman1}-\eqref{disuman2},  we therefore obtain the $G^{1}_{1}$ component of Einstein's tensor, $G^{\mu}_{\nu}\equiv R^{\mu}_{\nu}-\frac{1}{2}\delta^{\mu}_{\nu}R$, namely the pressure, as follows

\begin{equation}\label{circapressione}
P=-\frac{1}{8\pi G}\left[\frac{(a^\prime(\tau))^2}{a(\tau)^4}-2\frac{a^{\prime\prime}(\tau)}{a(\tau)^3}-2\frac{\Gamma_{,\tau\tau}}{a(\tau)^2}\right]\,.
\end{equation}

Immediately, assuming a matter dominated EdS universe after inflation, we find
\begin{subequations}
\begin{align}\label{pressure5000}
\Delta \rho&=0\,,\\
\Delta P&=-3H_I^2-2\Gamma_{,\tau\tau}\,,
\end{align}
\end{subequations}
where $\rho^{(1;2)}\equiv G_{0}^{0,(1;2)}$ and $P^{(1;2)}\equiv-\frac{1}{8\pi G} G_{1}^{1,(1;2)}$, \textit{i.e.}, density and pressure respectively whereas the density and pressure shifts are, by definition, $\Delta \rho\equiv\rho^{(1)}-\rho^{(2)}$ and $\Delta P\equiv P^{(1)}-P^{(2)}$. 

As a consequence, requiring $\Delta P\sim \rho_{cr}$ provides     

\begin{equation}
\Gamma_{,\tau\tau}\simeq \frac{\rho_{cr}-3H_I^2}{2}\,.
\end{equation}
Forcing $\Gamma_{,\tau\tau}$ to vanish suggests that $H_I^2\simeq \rho_{cr}$. Since this value is the today critical density previously introduced, it is quite likely that $H_I^2$ should be close to this value \emph{immediately after} inflation. This heuristic proof is supported by the fact that, although we denoted with $H_I$ the inflationary Hubble rate, it is not \emph{exactly} constant throughout inflation and in particular its value is much smaller than the one in Eq. \eqref{eq74} as inflation is ending.  Hence, its value, once the process of geometric particle production ends, is proportional to the current critical density as a consequence of our cancellation mechanism.

Accordingly, we infer that 
\begin{itemize}
    \item[--] on the surface $\tau=0$,  vacuum energy cancellation is associated to a minimum of the $\Gamma$ function, as $\rho_{cr}>3H_I^2$;
    \item[--] the total energy density is constant on $\tau=0$, \textit{i.e.}, $\rho^{(1)}=\rho^{(2)}$, implying energy conservation;
    \item[--] the pressure shift suggests that the corresponding fluid evolves as a \emph{dark fluid} \cite{dark}, mimicking the predictions presented in \cite{LM}.
\end{itemize}

\subsection{Consequences on  background cosmology}

As a consequence of our recipe, one  argues that the standard cosmological model, \textit{i.e.}, the $\Lambda$CDM paradigm, is modified because, at the end of our process, we can model the corresponding total fluid as a single fluid of matter whose pressure is not exactly zero, but is constrained to current value, called before $\Lambda_B$. Thus, the fine-tuning issue is no longer a real problem because quantum fluctuations associated to $\Lambda$ are removed by virtue of our cancellation mechanism. 

The value of $\Delta P$, however, is fixed at $\tau=0$. It is natural to wonder whether it remains constant throughout the evolution of the universe at late times, namely $\tau\rightarrow \infty$ or not. For the sake of simplicity, we may assume it to be constant without any time evolution for pressure, albeit we cannot exclude the pressure to vary at late-times. In the case of non-varying pressure, then the model reduces to the one presented in \cite{LM} with the great advantage to physically-explain how density,  cancelled out by the mechanism, transforms to new species of particles.

To evaluate $\Delta \rho$ and $\Delta P$ we made the ansatz of having a matter dominated EdS universe, characterized therefore by $P^{(2)}=0$. However, shifting to a radiation dominated EdS universe we again would get $\Gamma_{,\tau\tau}\simeq H_I^2$, implying that, at $\tau=0$, our model is not particularly influenced by choosing either matter or radiation. Then, by virtue of the continuity equation one computes a constant density that resembles the $\Lambda$CDM model, exhibiting a very different physical interpretation over the constant that fuels the universe to speed up today. 

This theoretical scheme works if $\Gamma$ is constant on the hypersurface $\tau=0$. Since inflation ends, requiring a perfect homogeneous and isotropic universe, one argues negligible $\Psi$ at the end of inflation, say $\Psi\rightarrow0$ as $\tau\rightarrow0$. From Eq. \eqref{ik}, assuming $\rho_{cr}-3H_I^2=\varepsilon\rho_{cr}$, we thus have

\begin{equation}\label{eq35bis2}
\ddot{\mathcal{F}}(\epsilon,0,H_I)=-\frac{\varepsilon}{2}\rho_{cr}\ e^{-i{\bf q}\cdot {\bf x}}\,,
\end{equation}

\noindent where $\varepsilon$ is a  unknown constant that quantifies the deviation between $H_I$ and $\rho_{cr}$. Afterwards, involving Eq. \eqref{eq54} and assuming the slow roll parameter to vanish after inflation in order to fulfill $\mathcal F= \dot{\mathcal{F}}=0$, we get  $\ddot{\mathcal{F}}=0$, implying $H_I^2=\rho_{cr}/3$, again addressing the coincidence problem\footnote{Inflation ends as $\epsilon\rightarrow1$. Thus, we justify the jump  to $\epsilon\simeq0$ noticing that the inflaton potential and vacuum energy disappear after inflation. So, both radiation and matter fields would dominate over any inflaton field. This permits one to presume that $\epsilon\simeq0$. }. Clearly, a more suitable choice of $\mathcal F$ is required to guarantee that $\mathcal F=\dot{\mathcal F}=0$ and $\ddot{\mathcal F} \neq0$ in general and this implies to select a more suitable version of the effective potential, instead of our hilltop quadratic one corrected by a Yukawa-like term involving a coupling with curvature.  

In general, however, addressing this issue is a central problem related to any  inflationary scenarios \cite{inflation}, whereas the here-employed potential only represents a first proposal to work out our model of dark matter production. The search for a more suitable version of the effective potential, however, requires to exit from inflation. This may be jeopardized by the coupling with curvature that, albeit it becomes negligibly small, is assumed to be small enough to guarantee $\epsilon\rightarrow1$ immediately before the jump to $\epsilon\rightarrow0$. Hence, a more suitable choice of the underlying potential would give new insights toward a graceful exit from inflation and at the same time the geometric production of dark matter particles. On the other side, we believe the need of curvature coupling is essential to interpret the corresponding dark matter fluid. In fact, if no coupling with curvature occurs, then the interpretation of particle production cannot be geometrical and only baryons can form during reheating as byproduct of the scalar field alone. We stressed such considerations throughout the text previously and we here underline that the more particles are produced from geometry the more coupling with $R$ is clearly needful, \textit{i.e.}, to fix the current dark matter abundance one has to invoke a further coupling.

\subsection{Dark matter with pressure?} \label{sezIVD}

In view of the aforementioned prescriptions, our corresponding dark energy scenario can be modeled using a dark fluid \cite{dunsby,LQ0}, effectively  compatible with the one presented in Refs. \cite{LQ,LQ2,LM,DLM}. This fluid can be interpreted as \emph{a single fluid of matter with pressure}, where the pressure is furnished by the additional bare cosmological constant. Indeed, if zero-point fluctuations are cancelled, leaving a $\Lambda_B\neq 0$, the remaining universe density would be associated to matter (and radiation, clearly), but the corresponding pressure would be given by the sum of $\Lambda_B$ and the pressure of dust and radiation. As the universe expands, radiation dominates over matter and $\Lambda_B$. But, since $\Lambda_B$ magnitude is comparable with matter, once the matter epoch finishes then $\Lambda_B$ tends to dominate over dark matter and baryons, reproducing \emph{de facto} the behavior of current cosmological model. Accordingly, we can quantify the pressure throughout the Universe evolution as follows:
\begin{itemize}
    \item[--] During inflation, our choice of the hilltop potential leads to vacuum energy ($\rho_{vac}$) domination, resulting in a large and negative pressure. This is a common trait to all inflationary models, and of course implies the violation of the strong energy condition \cite{referenza4}. In this phase, the corresponding universe dynamics is then described by a quasi de Sitter solution, whose deviations from the pure de Sitter are due to the inflaton fluctuations.
    \item[--] At the end of inflation, a large part of vacuum energy has been transformed into geometric particles, while the remaining contribution is responsible for ordinary baryonic production. The pressure associated to geometric particles ($-\Lambda_B$) has been computed in Sec. \ref{sezIVB} and corresponds to the transition from the inhomogeneous quasi-de Sitter phase to a matter-dominated one, which represents a well-known simplified scheme for reheating. Clearly, the strong energy condition is here restored.  
    \item[--] After reheating, we find the usual radiation and then matter eras. In such phases the total pressure is due to dust, radiation and the $\Lambda_B$ contribution. However, the latter is expected to be small compared with radiation at early stages, being compatible with the standard Big Bang model.
    \item[--] At late times, matter and radiation becomes subdominant with respect to the bare cosmological contribution, whose negative pressure is expected to drive the current Universe expansion. This implies a new violation of the strong energy condition, which however overcomes the coincidence and fine-tuning problems due to the geometric origin of such pressure, as previously discussed.
    \end{itemize}
In Tab. \ref{tab2} we summarize the phases described above, specifying the corresponding pressure due to the dominant fluid in each phase.
\begin{table}[h!]
\begin{center}
\begin{tabular}{p{2.2cm} p{3.4cm} p{2.6cm}} 
\hline \hline 
\rule{0pt}{3ex} & {\bf Phase} &  {\bf Pressure} \\
 \hline
 \rule{0pt}{3ex} $\tau_I < \tau < 0\ \ \ $  & Inflation$\ \ $   & $P \sim - \rho_{\rm vac} =- \Lambda^4$ \\ 
 \hline
 \rule{0pt}{3ex} $0 < \tau < \tau_r\ \ \ $ & Reheating$\ \ $ & $P \sim 0$\\
 \hline
 \rule{0pt}{3ex} $\tau_r < \tau < \tau_m\ \ \ $ & Radiation era$\ \ $ & $P \sim \rho_{\rm rad}(\tau)/3$ \\
 \hline
 \rule{0pt}{3ex} $\tau > \tau_m\ \ \ $ & Matter era$\ \ $ &  $P \sim 0$ \\
 \hline
 \rule{0pt}{3ex} $\tau \rightarrow +\infty\ \ \ $ & Bare CC domination$\ \ $ & $P \sim -\Lambda_B$ \\
 \hline \hline
\end{tabular}
\end{center}
\caption{Summary of the phases of the Universe evolution and corresponding value of the pressure in our model.} 
 \label{tab2}
\end{table}

Hence, the here-depicted overall paradigm  fully-degenerates with the $\Lambda$CDM model, being however physically  highly-different from it.

\section{Limits of our toy model and possible improvements} \label{sez3}

We below summarize some points that are crucial toward the understanding of how our toy model works.

\begin{itemize}
\item[--] We introduced a given momentum cut-off scale, which in our case is a consequence of lying on super-Hubble scales.  Super-Hubble scales are required in order to properly deal with the notion of particle. In other words, only after horizon crossing the inflaton fluctuations can be described classically, so that the number of particles could be in principle measured. The exact value of the cut-off is in principle arbitrary: in order to have a time-independent value, we selected $k,p < a_{\rm min}H_I/1000$, where $a_{\rm min} \equiv a(\tau_i)$.  This ensure that Eq. \eqref{eq64} is valid for all the modes considered in the interval $\tau_i < \tau<0$, thus allowing to evaluate numerically the integral \eqref{eq57}. However, as already noted, in this way we neglect the contribution due to modes which cross the horizon after $\tau_i$. A larger cut-off would result in a larger number of particles produced and, therefore, smaller values for the mass $m^*$. This does not appear as possible drawback of our paradigm, but rather a consequence of the scale factor and the inflationary potential invoked into calculations.

\item[--] Vacuum energy amount is not known \emph{a priori}. Again, this limitation is not related to our paradigm but rather on the scales used to quantify quantum fluctuations. We here selected Planck scales,  since we expect the inflationary universe to emerge from a quantum gravitational state, with an energy density comparable to Planck density  \cite{wein1,lombrisier}. However, standard model of particle physics scales \cite{copeland, mark, mark2}, namely electroweak and/or quantum chromodynamics scales, could also be investigated, in principle. In such cases, however, the Hubble rate would decrease a lot, and so it appears crucial the kind of energy scale we impose for $\Lambda$ in order to get both the number of dark matter particles produced through our mechanism and the field mode evolution throughout the investigated universe dynamics, as one sees from Eqs. \eqref{eq47} and \eqref{h1index}.

\item[--] In studying geometric  production, we neglected the role of back-reaction. As discussed in Sec. \ref{sezIIbk}, we expect back-reaction to damp out the initial perturbation as particles are produced, thus decreasing the particle production rate as $\tau \rightarrow 0$. This issue may be, at least partially, healed by increasing the total time interval in which geometric particle production may have taken place. However, a rigorous, and clearly numerical, treatment of back-reaction is required in future works in order to obtain a self-consistent study of gravitational particle production. A different dynamics for the perturbation potential $\Psi$ would also affect the exact value of the pressure shift introduced in Sec. \ref{sezIVB}, to justify the current value of the cosmological constant. 
\end{itemize}


\subsection{The role of quantum particle production} \label{sez3A}

Another feature of our model is to assume geometric particles to dominate over any quantum mechanisms of particle creations, as discussed in Sec. \ref{sez2}. This contribution is usually computed assuming the universe to expand from an asymptotically flat region (\emph{in}) to another one (\emph{out}), and computing the corresponding Bogoliubov coefficients for ladder operators. Details of such calculations are described in Appendix \ref{appB}. 

There, asymptotic flatness is required in order to properly define the notion of particle (and vacuum), which is not unique in curved spacetime. Including quantum particle production in our framework would imply:
\begin{itemize}
    \item[--] the production of particle-antiparticle pairs at zero and first geometric order \cite{par, referenza2}. So, as already noted in Sec. \ref{sez2B.3}, these particles can annihilate, without having enough time to significantly contribute to the net dark matter budget of the universe;
    \item[--] an additional contribution to the second-order number density, Eq. \eqref{ndens}, depending on the Bogoliubov coefficients $\beta_k$ and $\beta_p$, would enter dark matter production. This contribution is always positive (cfr. Eq. \eqref{B9}) and for this reason it would affect dark matter production increasing the total number density $N^{(2)}(0)$.
\end{itemize}
Concerning the last item above, we underline that when dealing with de Sitter spacetimes, the main conceptual problem of Eq. \eqref{eq43} is that it can describe an asymptotically flat universe only in remote past, $\tau \rightarrow - \infty$ but not around $\tau=0$. 

In other words, we do not have an asymptotically flat \emph{out} region. This issue is discussed in detail in \cite{referenza1}, where the authors show that if de Sitter spacetime is extended also to $0 < \tau < \infty$, asymptotic flatness is recovered at $\tau \rightarrow + \infty$ and no quantum particle production occurs\footnote{Such conclusion remains true even if spacetime passes through a coordinate singularity at $\tau=0$.}. Of course, this approach cannot be employed in realistic models of universe evolution, since it would neglect the EdS phases subsequent to inflation.

Alternatively, one could imagine that after the usual transition from inflation to radiation/matter domination, the universe finally reaches an adiabatic regime at late times, where the notion of particle becomes meaningful again \cite{damo, robb}. In \cite{robb} the authors show that in this framework “quantum" particle production from vacuum is non-negligible. This is true in particular for super-Hubble modes $k \ll a_{\rm end} H_I$ , where $a_{\rm end}$ is the scale factor at the end of inflation. They also notice that the particle abundance is larger if the inflationary energy scale is of the order of $10^{16}$ GeV, which fully agrees with the here-considered scales.

As discussed above, we thus expect that non-zero Bogoliubov coefficients will increase the total number density of particles produced at second geometric order. The possible inclusion of such “quantum" particle production in our framework will be subject of future works, albeit it is expected, in view of our above considerations, as a fraction of dark matter, rather than the main constituents.


\subsection{Comparing geometric particles with previous dark matter candidates}

A key assumption in our model is that dark matter arises as a geometric quasi-particle as a consequence of the coupling between inflaton field and spacetime curvature, without any further quantum couplings. So, our scenario \emph{does not} involve the concept of weakly interacting massive particles (WIMPs) as derived from effective extensions of the particle physics standard model. However, the physical meaning of our geometric particles shows very stable configurations  directly induced by gravity and, by construction, interacting with other objects only under the action of gravity/geometry. Better saying, we can figuring out sorts of weakly interacting geometric particles (WIGEP) that, differently from WIMPs \cite{wimps}, may have a collective behavior in making structures to form sharply in the very early universe, without extending the standard model of quantum field theory. These particles cannot form at more recent epochs, by virtue of the cosmological principle, \textit{i.e.}, when no perturbations are involved the WIGEP mechanism is suppressed. So, summing up, our WIGEP would 

\begin{itemize}
    \item[--] be stable immediately after the Big Bang, as \emph{a priori} they do not exhibit charges and thus they do not interact electromagnetically;
    \item[--] have been created in a very large amount as consequence of deleting out vacuum energy that transforms into geometric particles, due to inhomogeneities;
    \item[--] behave as collective particles in order to create enough overdensities capable of having galaxies as today we observe\footnote{Heuristically speaking, this is the main reason to deal with \emph{quasi-particle} behavior of those particles got from Yukawa interaction between inflaton and curvature.} by virtue of the perturbed spacetime involved into computation.
\end{itemize}

So, since our particles behave in a very similar way than previous expectations, despite non being WIMPs, it is possible to confront the kinds of interactions developed in previous literature, namely cosmologically-stable dark matter particles and SM particles. These models mainly consider spin-0, spin-1/2 and spin-1 dark matter candidates interacting with SM fields (mostly fermions) through spin-0 or spin-1 mediator fields, usually dubbed \emph{portals}\footnote{The interested reader can consult the review \cite{wimps} and references therein.}.

Among the \emph{plethora} of plausible portals, the simplest scenario is represented by the so-called  \emph{standard model portals}, in which dark matter interacts with the standard model of particle physics through the Higgs or the Z-boson. These portals have only two free parameters, \textit{i.e.}, the dark matter and portal mediator masses. For their simplicity, they are highly-predictive, albeit disfavored in some experimental limits \cite{escud}. The scheme which more closely resembles our approach includes the scalar Higgs boson. So, without assuming any CP-violation and taking into account a scalar field $\chi$ that describes dark matter, the corresponding interacting Lagrangian becomes
\be \label{portal}
\mathcal{L}_I=\xi_0 \lambda_\chi^H \chi^* \chi H^\dagger H,
\ee
where $H= \left( 0\ \frac{v_h+h}{\sqrt{2}}  \right)$ is the Higgs boson doublet, not to be confused with the Hubble rate, in  unitary gauge \cite{lanca}, $h$ the excited field, $v_h$ the ground state and $\xi_0=1/2\ (1)$ in case dark matter is (not) its own antiparticle. 

\begin{figure} [t!]
    \centering
    \includegraphics[scale=0.68]{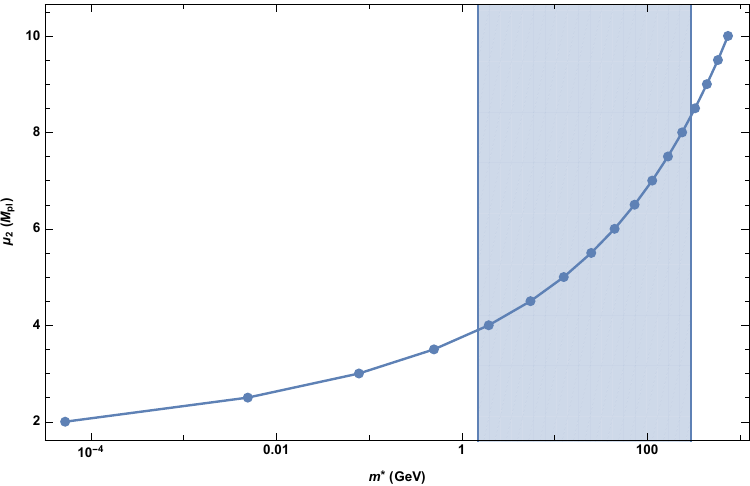}
    \caption{Allowed mass ranges for dark matter candidates assuming $\mu_2 \in [2,10]$. The other parameters are the same as in Fig. \ref{fig3}. The blue region is the excluded region got by the LUX limits for the Higgs portal, having set  $\lambda_\chi^H=1/6$ in analogy with our conformal coupling scenario. }
    \label{fig6}
\end{figure}

Confronting the two expectation limits for WIMPs and WIGEPs would therefore be extremely instructive to exclude mass ranges dropped out by observations. Hence, to do so we draw an excluding plot, see Fig. \ref{fig6}, where we show the mass value $m^*$ of our geometric dark matter candidate as function of the hilltop parameter $\mu_2$. There, a blue region that delimits the mass values that have been excluded by recent LUX limits \cite{wimps,lux} is prompted, as predicted by the above Higgs portal mechanism. Conventionally, for the sake of simplicity we have set $\xi_0=1$ and $\lambda_\chi^H=1/6$ in analogy with our scheme of field-curvature coupling\footnote{The effective potential of Eq. \eqref{effectivepotential} is clearly a Yukawa-like, whereas this is not. So, the choice of setting such constants in this way is only for a first naive confront between the two approaches. Changing the constant values, however, would not dramatically modify our conclusions. For instance, reducing $\lambda_\chi^H$ would get only narrower symmetric excluded regions.}. In case of large dark matter masses, we notice excluded regions lying inside

\begin{equation}\label{ancora}
1\,GeV \lesssim m^{*,excluded} \lesssim 400\, GeV\,, 
\end{equation}
where the superscript indicates the blue zone in Fig. \ref{fig6}. Consequently, by looking at our model, Eq. \eqref{ancora} would exclude hilltop $\mu_2$ values approximately inside the range 

\begin{equation}
\mu_{2}^{excluded} \in [4,8]\,. 
\end{equation}

We may therefore easily conclude that the WIGEPs may lie either in very large intervals of masses that are, somehow, similar to those predicted by the so-called WIMPzillas \cite{wimpzillas} or in smaller intervals, namely $\lesssim1$ GeV, being compatible with recent axion search \cite{axionsearch}.

It is remarkable to stress an important point as follows. Slow mass WIGEPs that we predict, or better to say that we cannot exclude, may be less likely since axions and/or in general ultralight fields may arise for other different processes, being dominant over geometric quasiparticles, see e.g. \cite{giusti1,giusti2,giusti3} where examples of virtual gravitons associated to the gravity sector can be found. These predictions look extremely similar to our outcomes, since we adopted a Yukawa-like interaction that couples the gravity and scalar field sectors adopting a fast interaction, similar to that discussed in the previous references. Toward our model,  since vacuum energy magnitude is huge, it is more likely that massive particles are produced, instead of light fields that would give a number of particles too large, i.e., far from expectations. Moreover, the more massive fields are predicted the more weakly interacting particles are expected, guaranteeing structures to form. We also stress that since dark matter is produced from the degrees of freedom coming from vacuum energy, but nothing has argued about the dynamics of the universe \emph{before} and \emph{after}, such  vacuum energy is associated to a symmetry breaking mechanism \cite{martin}. In this respect, it has been shown in Ref. \cite{LM,niki} that the solution of the cosmological constant problem would imply highly massive particles.

In other words, from the one side we \emph{do not} exclude ultralight fields to contribute to dark matter \cite{ultra}. On the other hand, however, we propose that the dominant contribution is geometrical, but with higher and more probable masses associated to it. 

In all the above treatment, particles have been produced assuming the free parameters to lie in suitable intervals that are compatible with theoretical expectations on vacuum energy and non-excluded regions provided by experiments. However, departures from these bounds may lead to different values of mass candidates, albeit the physical expectations about ultralight and highly-massive fields remain unaltered.

Phrasing it differently, we emphasize that, as characteristics of our proposed quasi-particles are better understood, it is conceivable that stricter allowable mass limits could be found, leaving the possibility that the mass range here discussed will be modified accordingly.

In addition to our analysis, it would be crucial to stress that the \emph{quantum cosmological constant problem} is not fully-addressed in our treatment. Indeed, even if we solve the classical cosmological
constant problem, finding a convincing reason to put the minimum of
the potential to zero, contributions from the zero-point fluctuations of all the
quantum fields present in the Universe cannot be ignored. They correspond to $\langle 0|T_{\mu\nu}|0\rangle$, with $T_{\mu\nu}$ the full energy momentum tensor, providing then a net contribution to the energy density that might be equated to the dark matter constituent. In our results, however, we did not assume the density provided by such a contribution. Clearly, further developments will focus on this central point.

\section{Final outlooks and perspectives}\label{sez4}

In this work, we proposed a toy model approach based on geometric cancellation of  vacuum energy, facing the classical cosmological constant problem.  To do so, we presumed that the corresponding large energy scales of $\Lambda$ transformed into geometric particles. In particular, in the primordial universe we assumed quantum fluctuations to be carried out by an effective hilltop potential, nonminimally coupled to scalar curvature $R$. So, without quantizing the fields, \textit{i.e.}, without assuming the quantum cosmological constant problem, we assumed a inhomogeneous quasi-de Sitter background universe, computing the corresponding particle production by means of perturbation theory in the external-field approximation. Once quantified the corresponding particle candidate, we interpreted it as dark matter, showing how the corresponding mass varies with respect to the free parameters of our model. We also showed that these mass limits are predicted to guarantee the cosmological constant contribution from vacuum energy is approximately canceled out. The remaining effective constant, namely the bare cosmological one, is therefore reviewed as responsible for the current acceleration, removing \emph{de facto} the fine-tuning issue. Our overall mechanism is prompted by requiring continuity of the universe and the validity of Israel matching conditions between the two inhomogeneous and homogeneous universes. This naturally shows how to generate geometric quasiparticles, that we conventionally called WIGEP in contrast to WIMPs. We provided a direct comparison among such particles and current bounds of dark matter masses, emphasizing excluded ranges and unsuited coupling constants that agree with present experimental windows. The model has been also confronted with previous literature and it has been argued that it guarantees a robust validity with respect to the paradigm developed in \cite{LM}, being compatible with predictions showed in \cite{DLM}. In this respect, a possible conjecture in which a dark fluid composed by a single dark matter fluid with pressure drives the universe today is also debated. 

Future works will refine the intervals of validity of our dark matter candidate, limiting the amount of mass associated to it. Moreover, we will investigate alternative versions of the effective inflationary potential that can quit inflation regardless the value of $R$, \textit{i.e.}, behaving as small fields after inflation and consequently check whether our predictions are particularly sensitive to the potential chosen for driving up cosmic inflation. 

Finally, we will investigate both the standard model of particle physics and the quantum cosmological constant problem, verifying whether our  here-presented toy model can be used even in case of field quantization and including the main features related to baryogenesis.

\acknowledgements
We express our gratitude to the unknown Referees for their fruitful comments that significantly helped to improve the quality of our manuscript.
O.L. thanks Simone Biondini, Rocco D'Agostino, Stefano Mancini, Marco Muccino, Hernando Quevedo, Alessandro Saltarelli and Sunny Vagnozzi for their suggestions and he is also grateful to Peter K. S. Dunsby for enlightening discussions on the topic of unified dark energy models made at the University of Cape Town. O.L. acknowledges the Ministry of Education and Science of the Republic of Kazakhstan, Grant: IRN AP08052311.


\newpage

\appendix

\vspace{3cm}

\section{Geometric particle production for a nonconformally coupled field} \label{appAA}

In this Appendix, we generalize our treatment assuming nonconformal coupling between inflaton and  scalar curvature. To do so, Eq. \eqref{eq3}  gives
\be \label{eqAA1}
\square_{\eta} \chi+ \left[ \left( \xi -\frac{1}{6} \right) R- \frac{2 \Lambda^4}{\mu_2^2} \right] a^2 \chi=0,
\ee
where $\chi$ is the rescaled field, introduced in Eq. \eqref{eq4}. 

Thus, explicitly computing the zeroth order scalar curvature 
\be \label{curvat}
R= 6 \ddot{a}/a^3= 12 H_I^2,
\ee
function of the exploited scale factor of Eq. \eqref{eq43}, and recalling the ansatz made in Eq. \eqref{eq9} for the modes, Eq. \eqref{eqAA1} becomes
\be \label{eqAA2}
\ddot{f}_k+ \left[ k^2- \frac{1}{(1-H_I \tau)^2}\left( \frac{2 \Lambda^4}{\mu_2^2}-12H_I^2\left(\xi - \frac{1}{6}\right) \right)  \right] f_k=0.
\ee
This equation admits a solution that resembles Eq. \eqref{eq47}, albeit  now the index of first kind Hankel's functions is 
\be \label{AA3}
\nu= \frac{9}{4}+ \frac{2 \Lambda^4}{\mu_2^2 H_I^2}-12 \xi,
\ee
which clearly reduces to Eq. \eqref{h1index} for conformal coupling. 

\begin{figure}[ht!]
    \centering
    \includegraphics[scale=0.6]{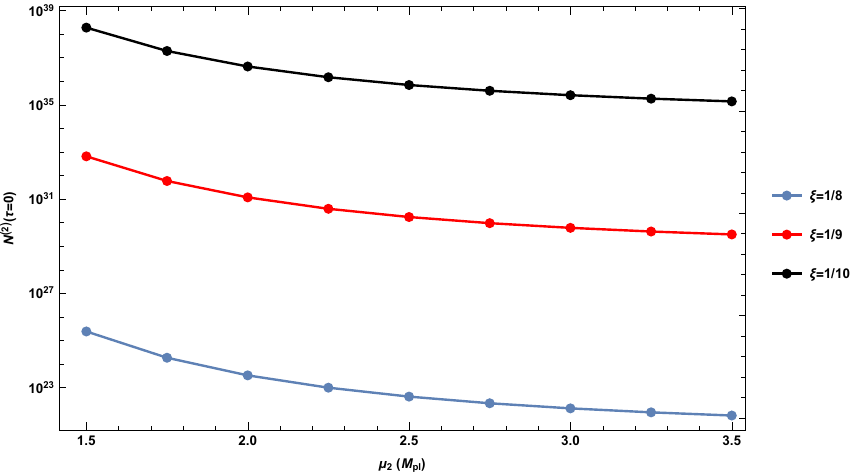}
    \caption{Number density $N^{(2)}$ in GeV$^3$ as function of the hilltop width $\mu_2$, in units of $M_{\rm pl}$. The other parameters are the same as in Fig. \ref{fig1}, with $\xi=1/C$ and $C=8;9;10$.}
    \label{fig1app}
\end{figure}

Consequently, in Fig. \ref{fig1app} we display how the number density of geometric particles changes by assuming three distinct cases for $\xi$, namely $\xi=1/C$ with $C=8;9;10$.

We notice that the number density of produced particles is larger for smaller coupling constants. Further, it is remarkable to stress that  larger densities are obtained for negative coupling. Hence, in order to obtain realistic values for the mass of the dark matter candidate, it is simply possible to modify the time $t_i$ at which geometric particle production is expected to begin, as discussed in Sec. \ref{sez2B.3}.


\section{Conformal and cosmic time} \label{appA}
Conformal time $\tau$ (or sometimes $\eta$) is related to cosmic time $t$ by
\be \label{eqA1}
d\tau= \frac{dt}{a},
\ee
where $a$ is the scale factor. A generic FRW line element in cosmic time, $ds^2=dt^2-a^2(t) d{\bf x}^2$, becomes $ds^2=a(\tau)^2\left[d\tau^2-d{\bf x}^2 \right]$, which is conformal to the Minkowski line element. Given a generic function of the cosmic time $f(t)$, the corresponding function in conformal time will be $f(\tau)$. For the latter, the following relations hold
\begin{align}
    &f^\prime(t)= \frac{\dot{f}(\tau)}{a(\tau)}, \label{eqA3} \\
    &f^{\prime \prime}(t)= \frac{\ddot{f}(\tau)}{a^2(\tau)}-\mathcal{H} \frac{\dot{f}(\tau)}{a^2(\tau)}, \label{eqA4}
\end{align}
where in our notation a prime denotes differentiation with respect to cosmic time and $\mathcal{H}=\dot{a}/a$. We have, in particular
\begin{align}
    & H= \frac{a^\prime}{a}= \frac{\dot{a}}{a^2}= \frac{\mathcal{H}}{a}, \label{eqA5} \\
    & a^{\prime \prime}= \frac{\ddot{a}}{a^2}- \frac{\mathcal{H}^2}{a}, \label{eqA6} \\
    & H^\prime= \frac{\dot{\mathcal{H}}}{a^2}- \frac{\mathcal{H}^2}{a^2}. \label{eqA7}
\end{align}
Easily the scale factor evolution in conformal time, in a EdS universe, can be obtained from Eq. \eqref{eqA1} to give
\be
a(t) \sim t^{2/(3+3w)}  \implies a(\tau) \sim \tau^{2/(1+3w)}\,,
\ee
where $w$ is a constant barotropic factor that provides the kind of EdS universe, e.g. for matter $w=0$, radiation $w={1\over 3}$, and so forth.

\section{Second order particle production in inhomogeneous spacetime} \label{appB}

As discussed in Sec. \ref{sez3A}, we assumed particle production  arising solely from the ``geometric'' contribution due to spacetime perturbations.

However, in a more general scenario, spacetime expansion can create particles independently from the presence of inhomogeneities, giving rise to  particle-antiparticle production from vacuum \cite{referenza1, par, for}. 

Moreover, it also implies additional contributions to the second order number density, prompted in Eq.  \eqref{ndens}, which we here discuss. 

So, assuming a spacetime with asymptotically flat \emph{in} and \emph{out} regions, we can introduce the \emph{Bogoliubov transformations}. To do so, we relate these quantities to  \emph{in} and \emph{out} ladder operators \cite{martin} 
\begin{align}
    & \hat{a}_{\text{out}}({\bf k})= \alpha_k^* \hat{a}_{\text{in}}({\bf k})- \beta^*_k \hat{a}^\dagger_{\text{in}}(-{\bf k}), \label{B1}  \\
    & \hat{a}_{\text{in}}({\bf k})= \alpha_k \hat{a}_{\text{out}}({\bf k})+ \beta^*_k \hat{a}^\dagger_{\text{out}}(-{\bf k}), \label{B2}
\end{align}
with $\alpha_k$, $\beta_k$ Bogoliubov coefficients. 

At first order in the inhomogeneities, the asymptotic \textit{out} state takes the form \cite{cesp}
\be \label{B3}
\lvert \Psi \rangle_{out}\equiv \lim_{\tau \rightarrow + \infty} \lvert \Psi \rangle = \mathcal{N} \left( \lvert 0, \text{in} \rangle + \frac{1}{2} S_{kp}^{(1)}\  \lvert {\bf k} {\bf p}, \text{in} \rangle  \right),
\ee
where $S_{kp}^{(1)}$ is a compact form for the probability amplitude given by Eq. \eqref{eq42}, whereas $\mathcal{N}=1+\mathcal O(h^2)$ is a normalization factor, arising from $\langle \Psi \rvert \Psi \rangle=1$. 

The number density in the \emph{out} region up to second order is then
\be \label{B4}
N= (2 \pi a)^{-3} \langle \Psi \lvert \int d^3q\  \hat{a}_{\rm out}({\bf q})^\dagger \hat{a}_{\rm out}({\bf q}) \rvert \Psi \rangle.
\ee
Besides normalization factors, the second order contribution reads
\begin{widetext}
\begin{align}
N^{(2)}&= \frac{1}{4}(2\pi a)^{-3} \int d^3k\ d^3p\  \lvert S^{(1)}_{kp}\rvert^2\ \langle {\bf k} {\bf p} \lvert  \hat{a}_{\rm out}^\dagger({\bf q}) \hat{a}_{\rm out}({\bf q}) \rvert {\bf k} {\bf p} \rangle \notag \\
&= \frac{1}{4} (2\pi a)^{-3} \int d^3k\ d^3p\   \lvert S^{(1)}_{kp}\rvert^2\ \langle {\bf k} {\bf p} \lvert \left(\alpha_q \hat{a}^\dagger_{\text{in}}({\bf q})- \beta_q \hat{a}_{\text{in}}(-{\bf q})\right)\left(\alpha_q^* \hat{a}_{\text{in}}({\bf q})- \beta^*_q \hat{a}^\dagger_{\text{in}}(-{\bf q})\right) \rvert {\bf k} {\bf p} \rangle, \label{B5}
\end{align}

where in the last step we have exploited the Bogoliubov transformation, Eq. \eqref{B1}. The only non-zero contribution to Eq. \eqref{B5} are
\be \label{B6}
N^{(2)}= \frac{1}{4} (2\pi a)^{-3} \int d^3k\ d^3p\ \lvert S^{(1)}_{kp}\rvert^2\ \langle {\bf k} {\bf p} \lvert \left(\lvert \alpha_q \rvert^2 \hat{a}_{\rm in}^\dagger({\bf q}) \hat{a}_{\rm in}({\bf q})+ \lvert \beta_q \rvert^2 \hat{a}_{\rm in}({\bf q})  \hat{a}_{\rm in}^\dagger({\bf q}) \right) \rvert {\bf k} {\bf p} \rangle.
\ee
In particular, Eq. \eqref{B6} gives
\be \label{B7}
N^{(2)}= \frac{1}{4} (2\pi a)^{-3} \int d^3k\ d^3p\ \lvert S^{(1)}_{kp}\rvert^2\ \left( \lvert \alpha_k \rvert^2 + \lvert \beta_k \rvert^2+ \lvert \alpha_p \rvert^2 + \lvert \beta_p \rvert^2 \right).
\ee
Exploiting now the  normalization condition for the Bogoliubov coefficients, namely $
\lvert \alpha_q \rvert^2-\lvert \beta_q \rvert^2=1$, with $q=k,p$, 
we finally infer
\be \label{B9}
N^{(2)}= \frac{1}{2} (2\pi a)^{-3} \int d^3k\ d^3p\ \lvert S^{(1)}_{kp}\rvert^2\ \left(\lvert \beta_k \rvert^2+ \lvert \beta_p \rvert^2 +1  \right),
\ee
which coincides with the result of Ref. \cite{referenza2}, up to a normalization factor. 
\end{widetext}
We notice then that second order number density contains a contribution which is independent from the relation between \emph{in} and \emph{out} vacua, \textit{i.e.}, the Bogoliubov coefficients, and becomes dominant in case of negligible $\beta_k$, $\beta_p$.

\end{document}